\documentclass[]{pasj01}
\draft

\begin{document} 
\Received{}%{yyyy/mm/dd}
\Accepted{}%{yyyy/mm/dd}
%\Published{yyyy/mm/dd}

\title{X-ray emissions from magnetic polar regions of neutron stars}

%%% begin:list of authors
% Do NOT capitalize all letters in "textsc".
\author{Hajime \textsc{Inoue}\altaffilmark{1}}%
%\thanks{Example: Present Address is xxxxxxxxxx}}
\altaffiltext{1}{ISAS, JAXA}
\email{inoue-ha@msc.biglobe.ne.jp}

%\author{B-Firstname \textsc{B-Familyname},\altaffilmark{2}}
%\altaffiltext{2}{B-Address of Institute}
%\email{bbbbb@xxx.xxx.xx.xx}

%\author{C-Firstname \textsc{C-Familyname}\altaffilmark{3}}
%\altaffiltext{3}{C-Address of Institute}
%\email{ccccc@xxx.xxx.xx.xx}
%%% end:list of authors

%% `\KeyWords{}' always has to be placed before `\maketitle'.
 %Do NOT move this preamble from here!

\maketitle

\begin{abstract}
Structures of X-ray emitting magnetic polar regions 
on neutron stars in X-ray pulsars are studied in a range of the accretion rate, 
10$^{17}$ g s$^{-1} \sim 10^{18}$ g s$^{-1}$.
It is shown that a thin but tall, radiation energy dominated, 
X-ray emitting polar cone appears at each of the polar regions.
The height of the polar cone is 
several times as large as the neutron star radius.
The energy gain due to the gravity of the neutron star in the polar cone 
exceeds the energy loss due to photon diffusion in the azimuthal direction of 
the cone, and a significant amount of energy is advected 
to the neutron star surface. 
Then, the radiation energy carried with the flow should become so large 
for the radiation pressure to overcome 
the magnetic pressure at the bottom of the cone.
As a result, the matter should expand 
in the tangential direction along the neutron star surface, 
dragging the magnetic lines of force, 
and form a mound-like structure.
The advected energy to the bottom of the cone should finally be radiated away from 
the surface of the polar mound and the matter should be settled 
on the neutron star surface there.
From such configurations, we can expect 
an X-ray spectrum composed of a multi-color blackbody spectrum 
from the polar cone region
and a quasi-single blackbody spectrum from the polar mound region. 
These spectral properties agree with observations.
A combination of a fairly sharp pencil beam and a broad fan beam is expected 
from the polar cone region, while a broad pencil beam is expected 
from the polar mound region.
With these X-ray beam properties, basic patterns of pulse profiles 
of X-ray pulsars can be explained too.

\end{abstract}

%%%%% {Introduction} %%%%%%%%%%%
\section{Introduction}\label{Introduction}
Soon after periodic X-ray variations were discovered from Cen X-3 
(Giacconi et al. 1971) and Her X-1 (Tananbaum et al. 1972), 
accretion environments of such sources were theoretically discussed 
by Pringle and Rees (1972); Lamb, Pethick and Pines (1973).
Since then, a large number of X-ray pulsars have been found and  
theoretical attempts have continuously been 
being done to explain the observational appearances of X-ray pulsars.
We have now the general consensus on situations of X-ray pulsars, settings of which 
could sequentially be summarized as follows,
\begin{itemize}
\item a close binary of a strongly magnetized neutron star + a ``normal" companion star,
\item an accretion flow from the companion star to the neutron star,
\item a magneto-boundary surface at $\sim 10^{8}$ cm from the neutron star where the magnetic pressure once stops the accretion flow,
\item two channeled flows along magnetic funnels towards the magnetic poles on 
the neutron star surface,
\item very hot regions at the bottoms of the channeled flows, after energy conversions 
of the kinetic energies to thermal energies through standing shocks,
\item X-ray emissions from the polar regions near the neutron star surface,
\item an oblique rotation of the magnetic axis, causing X-ray pulsations.
\end{itemize}

Pioneer studies on properties of the X-ray emitting regions 
were already done by some authors by the mid of 1970's.
Davidson (1973) proposed a situation 
that a hot, dense mound should be formed above each 
magnetic pole and that infalling material would be decelerated by the radiation pressure 
of photons trapped inside the mound.
He further argued that energy released above the mound should diffuse out as 
moderately hard X-rays and that energy released within the mound should emerge 
as soft X-rays from the whole surface of the neutron star.
By extending the idea of the radiation-pressure dominant polar mound to 
cases of high accretion rate, 
Inoue (1975) studied a structure of a radiation-pressure dominant polar cone
above each magnetic pole by taking account of the effect of the gravity there. 
He discussed that the gravitational energy released in the polar cone should be 
radiated away from the surface of the cone as a multi-color blackbody emission. 
Basco and Sunyaev (1975), on the other hand, discussed 
the generation and diffusion of radiation 
in a plasma with a strong magnetic field at the bottom of a magnetic funnel, 
supposing a fairly low accretion rate corresponding to 
the X-ray luminosity $\ll 10^{37}$ erg s$^{-1}$.

Even at this early moment in the mid of 1970's, a general picture on X-ray emission 
regions on each magnetic pole of X-ray pulsars was suggested.  
It was that there should exist   
two regions, the primary region where the kinematic energy of the infalling matter is 
converted the thermal energy and photons generated there or from the bottom side diffuse out upward suffering various 
interactions with matter in a magnetized plasma, 
and the secondary region where the accreted matter still gradually falls being braked by
the radiation pressure and the gravitational energy gained there finally diffuses out 
from its surface and/or the adjacent neutron star surface as blackbody emissions. 
Studies on the primary region have since been being done by several authors 
(e.g. Becker \& Wolff 2007; Wolff et al. 2016) but the secondary region has not been 
studied in more detail than done by Inoue (1975).

Davidson (1973) first introduced the secondary region as the dense mound 
at the base of the magnetic funnel.
This mound should be surrounded by the magnetic funnel and  
the infalling matter through the magnetic funnel is considered to finally accumulate there.
In the mound, a hydrostatic equilibrium should be established as
\begin{equation}
\frac{dP}{dz} = \rho \frac{GM}{R^{2}},
\label{eqn:MoundStr}
\end{equation}
where $P$ is the pressure at a position with a distance, $z$, from the mound 
bottom in the direction perpendicular to the stellar surface, 
$\rho$ is the matter density, $G$ is the gravitational constant, 
$M$ is the neutron star mass and $R$ is the neutron star radius.
Here, we have assumed the height of the mound, $h_{\rm M}$, is sufficiently 
smaller than the stellar radius.
Equation (\ref{eqn:MoundStr}) can be solved as
\begin{equation}
P_{*} = m \frac{GM}{R^{2}},
\label{eqn:P_*}
\end{equation}
where $P_{*}$ is the pressure at the bottom of the mound and that at the top has 
been approximated to be zero.
$m$ is the column density of the accreted matter in the mound defined as 
\begin{equation}
m = \int_{0}^{h} \rho dz.
\label{eqn:Def_m}
\end{equation}
The total mass of the mound, $M_{\rm M}$, is approximately given by 
introducing the average radius of the mound cross section, $x_{\rm M}$, as 
\begin{equation}
M_{\rm M} = \pi x_{\rm M}^{2} m.
\label{eqn:M_M-m}
\end{equation}
Let $t$ be time from the start of the accretion, then $M_{\rm M}$ should 
increase with $t$ as 
\begin{equation}
M_{\rm M} = \frac{\dot{M}}{2} \; t.  
\label{eqn:M_M-Mdot}
\end{equation}
$\dot{M}$ is the total accretion rate onto the neutron star 
and the factor (1/2) comes from an assumption that 
the flow should be equally divided into the two funnel flows to the respective 
N or S magnetic poles.
From these equations above, we see 
that the column density, $m$, should increase in time 
and the bottom pressure, $P_{*}$ should do so.
Then, at some moment, the pressure should overcome the magnetic pressure of 
the magnetic funnel and the matter should start to flow out 
from the mound bottom to the entire surface of the neutron star.
Thus, a steady state of the mass flow in the mound should be realized when 
the bottom pressure slightly exceeds the magnetic pressure.

The column density in the steady state, $m_{\rm M}$, is approximately gotten by 
equating the bottom pressure, $P_{*}$, to the bottom magnetic pressure, $(B^{2}/8\pi)_{*}$ in equation (\ref{eqn:P_*}) as
\begin{equation}
m_{\rm M} = \frac{(B^{2}/8\pi)_{*}} {GM/R^{2}},
\label{eqn:m_M}
\end{equation}
and the accumulation time of the matter in the mound, $t_{\rm A}$, is roughly 
estimated as
\begin{equation}
t_{\rm A} = \frac{m_{\rm M} \pi x_{\rm M}^{2}}{\dot{M} /2}.
\label{eqn:t_A}
\end{equation}
In the mound, the gravitational energy should still be released as the matter flows 
downwards and the energy generation rate should balance with the matter cooling rate 
in the steady state.
The matter cooling rate should be given by the photon diffusion 
in the azimuthal direction 
of the mound as far as $x_{\rm M} \ll h_{\rm M}$.
The side-way diffusion time, $t_{\rm D, M}$, is approximately given as
\begin{equation}
t_{\rm D, M} = \frac{3\kappa_{\rm T} \rho_{\rm M} x_{\rm M}^{2}}{4c}
\label{eqn:t_D}
\end{equation}
(see Inoue 1975), where $\kappa_{\rm T}$ is the Thomson scattering opacity,  and $c$ is the light velocity.  $\rho_{\rm M}$ is the average density of the mound and 
can be approximated as
\begin{equation}
\rho_{\rm M} \simeq \frac{m_{\rm M}}{h_{\rm M}}.
\label{eqn:rho_M}
\end{equation}
For the steady energy flow to realize, we set $t_{\rm A} = t_{\rm D}$ and get, 
from equations (\ref{eqn:t_A}), (\ref{eqn:t_D}) and (\ref{eqn:rho_M}),
\begin{equation}
h_{\rm M} \simeq \frac{3\kappa_{\rm T} }{8\pi c} \dot{M} = 1.4 \times 10^{5} \left( \frac{\dot{M}}{10^{17}\; \rm{g s}^{-1}} \right) \; \rm{cm}.
\label{eqn:h-Mdot}
\end{equation}
The above simple considerations indicate that the height of the mound, $h_{\rm M}$, 
could increase as the accretion rate, $\dot{M}$, increases 
and become comparable to or even larger than the stellar radius 
when $\dot{M}$ approaches to 10$^{18}$ g s$^{-1}$.
If $h_{\rm M}$ gets larger than $R$, X-ray luminosity from the secondary region could be dominant to that from the primary region,
since the energy release rate from the primary region and that from 
the secondary region are approximately given as 
($\dot{M}$/2)$GM/(R+h_{\rm M})$ and ($\dot{M}$/2)[$(GM/R)-(GM/(R+h_{\rm M}))$] respectively.

In this paper, we study nature and structure of the secondary region as a function of $\dot{M}$ more precisely than such simple arguments as done above.
Based on the results, then, we discuss 
observational appearances of X-ray emissions from there, 
focusing on cases of fairly high accretion rates corresponding to X-ray 
luminosities of 10$^{37} \sim 10^{38}$ erg s$^{-1}$.  
In section 2, the basic assumptions to study the accretion flow along the magnetic 
funnel are presented and structures of 
the X-ray emitting regions around the magnetic pole are solved through approximate 
equations. 
In section 3, we discuss observational appearances of X-ray spectra 
and pulse profiles expected from the magnetic polar regions as studied in section 2.
Summary and discussions are given in section 4.

%%%%% {Structures_magnetic_polar_regions} %%%%%%%%%%
\section{Structures of magnetic polar regions}

\subsection{Magneto-boundary surface}

Following the general consensus summarized in the previous section, 
we consider a situation that matter flows from a companion star into a region 
governed by gravitational field of a neutron star 
with magnetic field as strong as $10^{12} \sim 10^{13}$ gauss at its surface.
The magnetic field once halts the matter from further falling 
around the magneto-boundary surface at the distance $\sim 10^{8}$ cm 
from the neutron star.
The matter, then, flows along the magneto-boundary surface towards the magnetic 
poler regions and falls onto the neutron star through two funnel-like tubes guided 
by the magnetic lines of force.

The picture of the magneto-boundary surface was historically first studied 
in the case of roughly spherical accretion by a strongly magnetized compact star 
(e.g. Lamb, Pethick \& Pines 1973; Inoue \& H\={o}shi 1975; Arons \& Lea 1976;
Elsner \& Lamb 1977),
where the magnetic field is considered to be confined inside the magneto-spheric cavity because of complete screening of the stellar magnetic field from the outside plasma by currents in the transition region between the magneto-boundary surface and the plasma.

In the case that a thin disk extends down to the magneto-bounding surface, 
which should be more realistic in most of X-ray pulsars than the spherical flow, 
the accretion flow from the disk to the neutron star surface was studied in detail 
by Gosh, Lamb \& Pethick (1977); Gosh \& Lamb (1978, 1979).
It is shown that the stellar magnetic field penetrates the inner part of the disk 
through the Kelvin-Helmholtz instability, turbulent diffusion and so on, and that 
a transition zone arises in which magnetic coupling between the star and the disk 
transfers the angular momentum from the disk to the star.
The accreted matter is, then, considered to flow from the innermost region of the disk 
along the lines of force of the stellar magnetic field onto the polar regions of the stellar 
surface by getting over the barrier of the centrifugal force through  
the angular momentum transfer.

\begin{figure}
\begin{center}
\includegraphics[width=12cm]{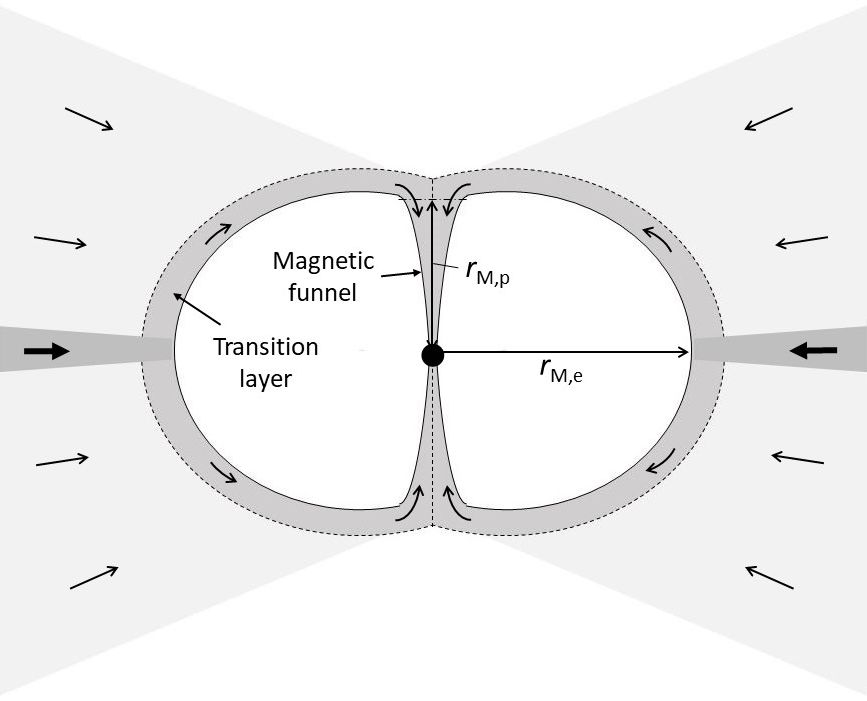} 
\caption{Schematic cross section of magneto-boundary surface considered in this paper. } \label{MagnetoBoundarySurface}
\end{center}
\end{figure}

In the scheme of Gosh \& Lamb (1978; 1979), only a thin disk is assumed 
to extend down to the magneto-boundary region and $\sim$ 20\% of 
the stellar magnetic field is discussed to remain outside the inner edge of the disk.
Recent observations, however, indicate that a geometrically thick flow could exist 
in parallel to the thin disk from the outermost part of the accretion disk 
(Churazov et al. 2001;Sugimoto et al. 2016; Inoue in preparation). 
If so, the thick flow should completely screen the remaining magnetic field and thus 
such a configuration as in figure \ref{MagnetoBoundarySurface} could be expected.
As discussed by Ghosh \& Lamb (1978; 1979), a fraction of the magnetic field should penetrate the inner part of the thin disk and the matter flows along the lines of force 
of the filed threaded in the disk 
by losing its angular momentum through the magnetic stress force.  
A small fraction of the magnetic field could remain outside the transition layer 
in which the matter is flowing from the thin disk but should completely be compressed 
and screened by the plasma in the thick disk.
The matter in the thick disk could go into the transition layer through instabilities 
such as the Reyleigh-Taylor instability (Arons \& Lea 1976; Elsner \& Lamb 1977), 
and flow together with the matter from the thick disk onto the neutron star surface.
As a result, the outer surface of the transition layer could have a shape similar to 
those obtained by Arons \& Lea (1976); Elsner \& Lamb (1977), as schematically drawn 
in figure \ref{MagnetoBoundarySurface}, where two cusps appear on the magnetic axis.
Here, we assume for simplicity that 
the magnetic axis and the rotational axis of the neutron star are aligned with each other 
and are perpendicular to the disk plane, as done in the above literatures.

%%%%% figure {MagnetoBoundarySurface}

Let $r_{\rm M, e}$ and $r_{\rm M, p}$ be 
the distances of the inner surface of the transition layer 
in the magneto-boundary surface on the equatorial plane 
and on the magnetic axis, respectively.
We assume 
\begin{equation}
r_{\rm{M,e}} = 0.4 \; r_{\rm{M,0}}, 
\label{eqn_Rel_r_p-r_e}
\end{equation}
adopting the result from Gosh \& Lamb (1978), where
$r_{\rm{M,0}}$ is the characteristic radius of the magneto-boundary surface for spherical 
accretion, and 
\begin{eqnarray}
r_{\rm{M,p}} &=& 0.5 \; r_{\rm {M,e}} \nonumber \\ &\simeq& 6.4 \times 10^{7} \left(\frac{\dot{M}}{10^{17} \; \rm{g \; s}^{-1}} \right)^{-2/7} \left(\frac{M}{M_{\odot}} \right)^{-1/7} \left( \frac{\mu_{\rm M}}{10^{30} \; \rm{gauss \; cm^{3}}} \right)^{4/7},
\label{eqn:r_Me}
\end{eqnarray}
adopting the value by Arons \& Lea (1976).
$\mu_{\rm M}$ is the magnetic moment of the neutron star.
Considering a situation that the accreted matter with the density, $\rho_{\rm T}$, falls with 
the velocity, $v_{\rm T}$, in the transition layer with the thickness, $h_{\rm T}$, from 
the equatorial side to the polar side, the mass continuity is approximately expressed as 
\begin{equation}
\frac{\dot{M}}{2} \simeq \rho_{\rm T} v_{\rm T} h_{\rm T} 2\pi r \sin \theta,
\label{eqn:CntEq_TL}
\end{equation}
where the factor of 1/2 comes from the presence of the two separate streams toward 
the two respective poles, and $\theta$ is the tangential angle of a relevant position 
from the polar axis.
Here and hereafter, we assume that the identical things happen in both sides of the magnetic poles.

The pressure of the matter in the transition layer, $P_{\rm T}$, 
is expressed with help of equation (\ref{eqn:CntEq_TL}) as 
\begin{eqnarray}
P_{\rm T} &=& \frac{2 \rho_{\rm T}}{m_{\rm p}} kT_{\rm T} \nonumber \\ &\simeq& \frac{\dot{M} kT_{\rm T}}{m_{\rm p}v_{\rm T} h_{\rm T} 2 \pi r \sin \theta},
\label{eqn:P_TL}
\end{eqnarray}
where $m_{p}$ is the proton mass, $k$ is the Boltzmann constant and $T_{\rm T}$ 
is the temperature of the flowing matter. 
The flow velocity, $v_{\rm T}$, should be given as a result of the angular momentum 
transfer in the transition layer and thus be proportional to the Alfven velocity there. 
In the present situation, the Alfven velocity should roughly equal to the sound velocity 
and hence $v_{\rm T}$ should be constant if the temperature is constant throughout 
the transition layer.
Under the isothermal approximation, $P_{\rm T}$ can be written in terms of that around the equatorial plane, $P_{\rm T,e}$ as
\begin{equation}
P_{\rm T} \simeq P_{\rm T,e} \left( \frac{r}{r_{\rm M,e}} \right)^{-1} \sin^{-1} \theta.
\label{eqn:P_M} 
\end{equation}
The magnetic pressure in the transition layer, $P_{\rm B}$, 
is, on the other hand, governed 
mainly by the magnetic field in the $\theta$ direction, 
which could be proportional to $r^{-3} \sin \theta$.
Hence,
$P_{\rm B}$ is approximately given, referring to the value around the equatorial plane, $P_{\rm B, e}$, as
\begin{equation}
P_{\rm B} \simeq P_{\rm B,e} \left(\frac{r}{r_{\rm M,e}}\right)^{-6} \sin^{2} \theta.
\label{eqn:P_B}
\end{equation}
Since $P_{\rm T} \simeq P_{\rm B}$ and $P_{\rm T,e} \simeq P_{\rm B,e}$, 
we get an approximate relation between $\theta$ and $r$ as
\begin{equation}
\sin \theta \simeq \left( \frac{r}{r_{\rm M,e}} \right)^{5/3}.
\label{eqn:Rel_theta-r}
\end{equation}
From this relation, we can see that there should be no solution near the cusp, 
where $r \simeq r_{\rm M,p}$ and $\theta < 1$, for the magnetic pressure 
of the magneto-boundary surface to balance with the matter pressure.
The opening angle, $\Theta_{0}$, of the region, in which the matter tends 
to radially fall onto the stellar surface without halted by the magnetic pressure, 
is approximately calculated from equation (\ref{eqn:Rel_theta-r}) with help of equation (\ref{eqn:r_Me}) as 
\begin{eqnarray}
\Theta_{0} &\simeq& \left(\frac{r_{\rm M,p}}{r_{\rm M,e}}\right)^{5/3} \nonumber \\
&\simeq& 0.32.
\label{eqn:Theta_0}
\end{eqnarray}

We assume that the magnetic field tends to be dipole-like and the matter should infall to the stellar surface along the lines of force in this region with 
$\theta \leq \Theta_{0}$, and call this region as the magnetic funnel, hereafter.
We, then, approximate the opening angle of the magnetic funnel, $\Theta$, to be a function of $r$, as 
\begin{equation}
\Theta = \Theta_{0} \left(\frac{r}{r_{\rm M,p}} \right)^{1/2},
\label{eqn:Theta}
\end{equation}
guided by the dipole-field lines of force.
This equation can be rewritten with help of equations (\ref{eqn_Rel_r_p-r_e}),  
(\ref{eqn:r_Me})  and (\ref{eqn:Theta_0}) as
\begin{equation}
\Theta = \Theta_{*} \left(\frac{r}{R} \right)^{1/2},
\label{eqn:Theta_r/R}
\end{equation}
where $\Theta_{*}$ is the opening angle of the funnel on the surface of the neutron star 
and is as
\begin{equation}
\Theta_{*} = 4.0 \times 10^{-2} \left(\frac{R}{10^{6} \; \rm{cm}} \right)^{1/2} \left(\frac{\dot{M}}{10^{17} \; \rm{g \; s}^{-1}} \right)^{1/7} \left(\frac{M}{M_{\odot}} \right)^{1/14} \left( \frac{\mu_{\rm M}}{10^{30} \; \rm{gauss \; cm^{3}}} \right)^{-2/7}.
\label{eqn:Theta_*}
\end{equation}

\subsection{Overall picture of the magnetic polar regions}
Since the matter falling in the magnetic funnel, 
finally hits the neutron star surface, a standing shock should appear 
at a certain height from the stellar surface.
The matter can be approximated to flow with the free fall velocity 
on the upper-stream side of the shock.  
We call this region as the free fall region. 

The kinetic energy of the inflowing matter should be converted to the thermal 
energy through the shock.  Thermal emissions are expected mainly in the X-ray band 
from the region between the shock and the stellar surface.  
We call this X-ray emitting region as the polar cone.
The thermal pressure is lower than the magnetic pressure in the polar cone, 
and the matter falls inwards along the magnetic lines of force there.

At the bottom of the polar cone, however, the matter pressure comes to exceed the magnetic 
pressure and the matter flows out along the stellar surface from the polar cone 
by dragging the magnetic lines of force.
A mound-like structure appears at each of the two polar surface regions and 
additional X-ray emissions are expected from there.  We call these mound-like structures as the polar mounds hereafter.

\begin{figure}
\begin{center}
\includegraphics[width=8cm]{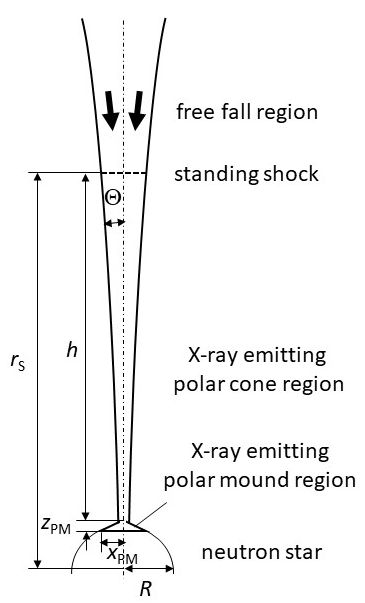} 
\caption{Schematic diagram of magnetic polar regions considered in this paper. } \label{MagneticPolarRegions}
\end{center}
\end{figure}

Figure \ref{MagneticPolarRegions} shows the schematic diagram of the magnetic polar regions, considered in this paper.

%%%%% figure {MagneticPolarRegions}

\subsection{Free fall region}

The matter is assumed to flow with the free fall velocity in the free fall region 
and then the radial inflow velocity, $v_{\rm F}$, 
at a position with a distance, $r$, from the center of the neutron star is given as
\begin{equation}
v_{\rm F} = \left(\frac{2GM}{r} \right)^{1/2}.
\label{eqn:v_r}
\end{equation}
The continuity equation is written as
\begin{equation}
\rho_{\rm F} v_{\rm r} \pi  (r \Theta)^{2} = \frac{\dot{M}}{2}, 
\label{eqn:ContEq}
\end{equation}
where $\rho_{\rm F}$ is the density of the matter in the free fall region and is gotten as
\begin{equation}
\rho_{\rm F} = \frac{\dot{M}}{2\pi \Theta_{*}^{2} (2GM)^{1/2} R^{3/2}} \left(\frac{r}{R} \right)^{-5/2}.
\label{eqn:rho_F}
\end{equation}

The optical depth for Thomson scatterings, $\tau_{\rm F}$, in the azimuthal direction 
in the free fall region 
is calculated with equations (\ref{eqn:Theta_r/R}) and (\ref{eqn:rho_F}) as
\begin{eqnarray}
\tau_{\rm F} &=& \kappa_{\rm T} \rho_{\rm F} r \Theta  \nonumber \\ &\simeq& 9.9 \left( \frac{\dot{M}}{10^{17} \, \rm{ g \, s}^{-1}} \right)^{6/7} \left(\frac{M}{M_{\odot}} \right)^{-4/7} \left(\frac{\mu_{\rm M}}{10^{30} \, \rm{gauss \, cm}^{3}} \right)^{-2/7}  \left( \frac{r}{10^{6} \rm{cm}} \right)^{-1},
\label{eqn:tau}
\end{eqnarray}
where $\kappa_{\rm T}$ is the opacity of the Thomson scattering.
From this equation, we see that the flow within the magnetic funnel 
above the shock front is optical thick for the Thomson scattering 
in case of $\dot{M} \gtrsim 10^{17}$ g s$^{-1}$ unless $r \gg 10^{7}$ cm.

%%%%% subsection {XEPC} %%%%%%%%%%
\subsection{X-ray emitting polar cone region}\label{XEPC}

%%%%% subsubsection{RadEnDominance}
\subsubsection{Dominance of the radiation energy in the polar cone} \label{RadEnDominance}
As shown in equation (\ref{eqn:tau}), the optical depth is larger than unity 
even in the free fall region in front of the shock.
The density behind the shock front must be larger than the density in the free fall region 
and the density gets larger and 
larger as the matter in-falls in the polar cone.
Thus, we can say that polar cone is sufficiently optically thick.

The density behind the shock, $\rho_{\rm S}$, is given as
\begin{equation}
\rho_{\rm S} = \frac{\gamma+1}{\gamma-1} \rho_{\rm F} (r_{\rm S}),
\label{eqn:rhoS}
\end{equation}
where $\gamma$ is the specific heat ratio.
If we initially assume that the gaseous energy is dominant to the radiation energy, 
$\gamma$ should be 5/3 and then $\rho_{\rm S} = 4 \rho_{\rm F} (r_{\rm S})$.
The ion temperature behind the shock, $T_{\rm i, S}$, is given as
\begin{equation}
T_{\rm i, S} = \frac{3}{8} \frac{m_{\rm p}}{k} \frac{GM}{ r_{\rm S}}.
\label{eqn:T_iS}
\end{equation}
The electron temperature, $T_{\rm e, S}$ should be determined by a balance 
between energy input through collisions with protons and energy loss 
through photon emissions and Compton scatterings.  
The energy input rate per mass, $\varepsilon_{+}$, is expressed as
\begin{equation}
\dot{\varepsilon}_{+} = \frac{3 kT_{\rm i,S}/ 2m_{\rm p}} {t_{\rm ie}},
\label{eqn:epsilon_+}
\end{equation}
when $T_{\rm i,S} \gg T_{\rm e,S}$.
$t_{\rm ie}$ is the ion-electron equipartition time and is given by Spitzer (1962) as 
\begin{equation}
t_{\rm ie} = 4.2 \times 10^{-22} (\ln \Lambda)^{-1} \rho_{\rm S}^{-1} T_{\rm e,S}^{3/2}
\label{eqn:t_ie}
\end{equation}
in cgs unit, where $\ln \Lambda$ is the Coulomb logarithm.
The energy loss rate per mass, $\dot{\varepsilon}_{-}$ is, on the other hand, written as 
\begin{equation}
\dot{\varepsilon}_{-} = A \dot{\varepsilon}_{0} \rho_{\rm S} T_{\rm e, S}^{1/2},
\label{eqn:epsilon_-}
\end{equation}
where  $\dot{\varepsilon}_{0} \rho_{\rm S} T_{\rm e, S}^{1/2}$ is the free-free emission 
rate per mass and $A$ is an amplification factor through 
inverse Compton scatterings. 
From $\dot{\varepsilon}_{+} = \dot{\varepsilon}_{-}$ with help of equations (\ref{eqn:T_iS}) $\sim$ (\ref{eqn:epsilon_-}), we get, independently of the density, 
\begin{equation}
T_{\rm e,S} = 1.9 \times 10^{10} \left(\frac{M}{M_{\odot}}\right)^{1/2} \left(\frac{r_{\rm S}}{10^{6} \; \rm cm}\right)^{-1/2} \rm K,
\label{eqn:T_eS}
\end{equation}
where we have assumed $\ln \Lambda = 10$ and $A = 10$ (Illarionov and Sunyaev 1972).

Since $\dot{\varepsilon}_{-}$ is thought to be the radiation energy generation rate per mass, 
we can approximately calculate the radiation energy density per mass, 
$\varepsilon_{\rm Rad,S}$, 
expected when the electron temperature is as in equation (\ref{eqn:T_eS}), as
\begin{equation}
\varepsilon_{\rm Rad,S} \simeq \dot{\varepsilon}_{-} \; t_{\rm D,S}.
\label{eqn:Def_epsilon_Rad}
\end{equation}
Here, $t_{\rm D, S}$ is a time scale, during which the generated photons diffuse out 
in the azimuthal direction of the polar cone, and is approximately represented as 
\begin{equation}
t_{\rm D, S} = \frac{3 \kappa_{\rm T} \rho_{\rm S} (r_{\rm S} \Theta_{\rm S})^{2}}{4 c}.
\label{eqn:tDS}
\end{equation}

If $\varepsilon_{\rm Rad, S}$ is as large as or larger than the gaseous energy per mass behind the shock, which is given as 
\begin{equation}
\varepsilon_{\rm Gas, S} = \frac{9}{16} \frac{GM}{r_{\rm S}},
\label{eqn:epsilon_GasS}
\end{equation}
with the specific heat ratio of 5/3, 
we can assure that the radiation energy is sufficiently established to be dominant to the gaseous energy in the polar cone.

\begin{figure}
\begin{center}
\includegraphics[width=8cm]{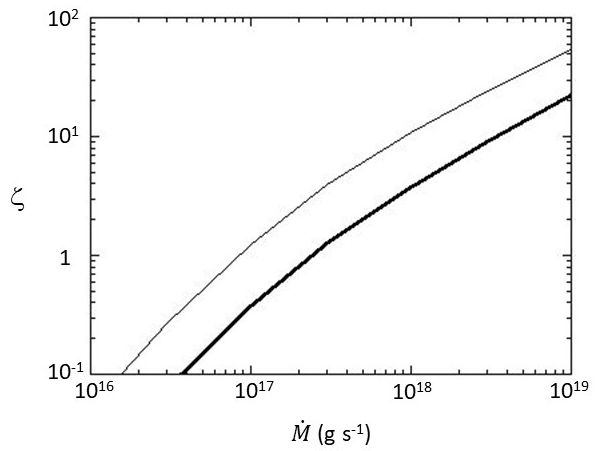} 
\caption{$\zeta$-value, which indicate the dominance of the radiation energy density 
behind the shock, as a function of $\dot{M}$.
The thick and thin lines correspond to cases of $\mu_{\rm M} = 10^{30}$ gauss cm$^{3}$ and $\mu_{\rm M} = 10^{31}$ gauss cm$^{3}$ respectively.}\label{Mdot-zeta}
\end{center}
\end{figure}

Let $\zeta$ be the ratio of $\varepsilon_{\rm Rad, S}$ to $\varepsilon_{\rm Gas, S}$, 
and it is given from above equations as
\begin{eqnarray}
\zeta &=& \frac{\varepsilon_{\rm Rad,S}}{\varepsilon_{\rm Gas,S}} \nonumber \\
&=& 0.9 \,  \left(\frac{\dot{M}}{10^{17} \rm \, g \, s^{-1}} \right)^{12/7} \left(\frac{\mu_{\rm M}}{10^{30} \rm \,gauss \, cm^{3}} \right)^{4/7} \left(\frac{M}{M_{\odot}} \right)^{-53/28} \left(\frac{r_{\rm S}}{10^{6} \rm{\, cm}} \right)^{-5/4},
\label{eqn:zeta}
\end{eqnarray} 
where we have assumed $A=10$ and $\ln \Lambda = 10$ again.
By using the $r_{\rm S}$ and $\dot{M}$ relation for the polar cone as given in figure  \ref{Mdot-Height_Relations} in \ref{Structure}, 
we can calculate $\zeta$ as a function of $\dot{M}$ for two cases of $\mu_{\rm M}$ and the result is shown in 
figure \ref{Mdot-zeta}.
We see that $\zeta \gtrsim 1$, when $\dot{M} \gtrsim 10^{17}$ g s$^{-1}$.
Since our concern mainly exists on the bright X-ray pulsars with $\dot{M} \gtrsim 10^{17}$ g s$^{-1}$ here, we approximate the thermal energy as being governed 
by the radiation energy, and neglect the gaseous energy in the polar cone.

%%%%% figure {Mdot-zeta} %%%%%

\subsubsection{Energy loss due to side-say diffusion of photons}
Since $\Theta \ll 1$, photons in the polar cone tend to diffuse out 
in the azimuthal direction of the cone.
Thus, we approximate that the energy loss of the inflowing matter in the polar cone 
is governed by the side-way diffusion of photons.
In that case, the energy loss rate per unit volume, $q$, is approximately given as
\begin{equation}
q = \frac{u}{t_{\rm D}},
\label{eqn:Def_q}
\end{equation}
where $u$ is the typical radiative energy density in the polar cone, 
and $t_{\rm D}$ is the diffusion time as in equation (\ref{eqn:tDS}).
The equation for the energy flow in the polar cone is then given as
\begin{eqnarray}
\frac{\dot{M}}{2} \frac{d}{dr}\left(\frac{v^{2}}{2} + \frac{4}{3} \varepsilon - \frac{GM_{\rm X}}{r} \right) &=& \pi (r \Theta)^{2} q \nonumber \\
&=& \frac{4 \pi c}{3 \kappa_{\rm T}} \varepsilon,
\label{eqn:BernoulliEq}
\end{eqnarray}
with help of equation (\ref{eqn:Def_q}), equation $u = \rho \varepsilon$, and equation (\ref{eqn:tDS}), 
where $v$ and $\varepsilon$ are respectively the inflow velocity and 
the radiative energy per unit mass in the polar cone.

%%%%% subsubsection {Structure}
\subsubsection{Structure of the polar cone}
\label{Structure}

Based on the above arguments, equations determining structures of the polar cone are
as follows.

The first is the continuity equation which is given in equation (\ref{eqn:ContEq}) 
by replacing $v_{\rm T}$ with $v$.
Here, we assume for simplicity 
that $\rho$ and $v$ are uniform on a cross section 
in the azimuthal direction of the polar cone.

The second is the dynamical equation, 
on an assumption that the flow in the polar cone is sufficiently subsonic 
to neglect the velocity gradient term,
\begin{equation}
\frac{dP}{dr} = - \rho \frac{GM}{r^{2}}.
\label{eqn:DB_Eq}
\end{equation}

The third is the energy equation, modified from equation (\ref{eqn:BernoulliEq}) by neglecting the kinetic energy of the matter flow, as
\begin{equation}
\frac{d}{dr} (\frac{4}{3} \varepsilon - \frac{GM}{r}) = \frac{\varepsilon}{r_{\rm D}},
\label{eqn:EnEq}
\end{equation}
where $r_{\rm D}$ is a parameter 
defined as
\begin{equation}
r_{\rm D} = \frac{3 \kappa_{\rm T} \dot{M}}{8 \pi c} = 1.4 \times 10^{5} \left(\frac{\dot{M}}{10^{17} \rm{g \; s}^{-1}} \right) \; \rm{cm}.
\label{eqn:Def_rD}
\end{equation}

In the above two equations, we have assumed 
that $P$ and $\varepsilon$ are radiation pressure and radiation energy 
per mass, neglecting gaseous pressure and energy, 
and that both are representative quantities over a cross section of the polar cone.
Then, we can set
\begin{equation}
P = \frac{u}{3} = \frac{\rho \varepsilon}{3}.
\label{eqn:EqState}
\end{equation}

From equations (\ref{eqn:DB_Eq}), (\ref{eqn:EnEq}) and (\ref{eqn:EqState}), we get 
two equations to determine $\varepsilon$ and $\rho$ as functions of $r$ as
\begin{equation}
\frac{d\varepsilon}{dr} = \frac{3}{4} ( \frac{\varepsilon}{r_{\rm D}} - \frac{GM}{r^{2}} )、
\label{eqn:DifEq1}
\end{equation}
and
\begin{equation}
\frac{d\rho}{dr} = -\frac{3\rho}{4\varepsilon} ( \frac{\varepsilon}{r_{\rm D}} + 3  \frac{GM}{r^{2}} ).
\label{eqn:DifEq2}
\end{equation}

The top boundary of polar cone is the shock front.
The density at the top boundary should be given by equation (\ref{eqn:rhoS}).
Since the radiation energy density is considered to be dominant in the polar cone, 
we set $\gamma = 4/3$ in the polar cone and then $\rho_{\rm S} = 7 \rho_{\rm F} (r_{\rm S})$.
%Although there could be arguments on the shock structure (e.g. Blandford \& Payne 1981), it should be noted that the solutions obtained below do not depend on 
%this boundary condition largely.
By also approximating that the kinetic energy of the inflow matter 
in front of the shock is all converted to the radiative energy (enthalpy) behind 
the shock, $\varepsilon_{\rm S}$ is expressed as
\begin{equation}
\varepsilon_{\rm S} = \frac{3}{4} \frac{GM}{r_{\rm S}}.
\label{eqn:OuterBC}
\end{equation}

These boundary values are functions of $r_{\rm S}$, and thus we numerically solve 
equations (\ref{eqn:DifEq1}) and (\ref{eqn:DifEq2}) from an assumed position, 
$r_{\rm S}$, successively decreasing $r$ inward to the bottom boundary 
at the neutron star surface where $r = R$.
As shown later, the radiative pressure steeply increases as $r$ decreases 
and comes to exceed the magnetic pressure of the funnel at some position.
The magnetic funnel should be pushed out, when the radiative pressure exceeds 
the magnetic pressure, and the matter should flow out along the neutron star surface 
by dragging the magnetic lines of force from the funnel-like magnetic column.
Thus, the following condition should be satisfied at the bottom of the polar cone as 
\begin{equation}
P_{*} = \frac{\rho_{*} \varepsilon_{*}}{3} = \frac{B_{r, *}^{2}}{8 \pi},
\label{eqn:BtmBC}
\end{equation}
where we represent 
the physical quantities at the bottom of the polar cone with the subscript *.
The distance of the top boundary of the polar cone, $r_{\rm S}$, 
has been searched to satisfy this bottom boundary condition.

\begin{figure}
\begin{center}
\includegraphics[width=15cm]{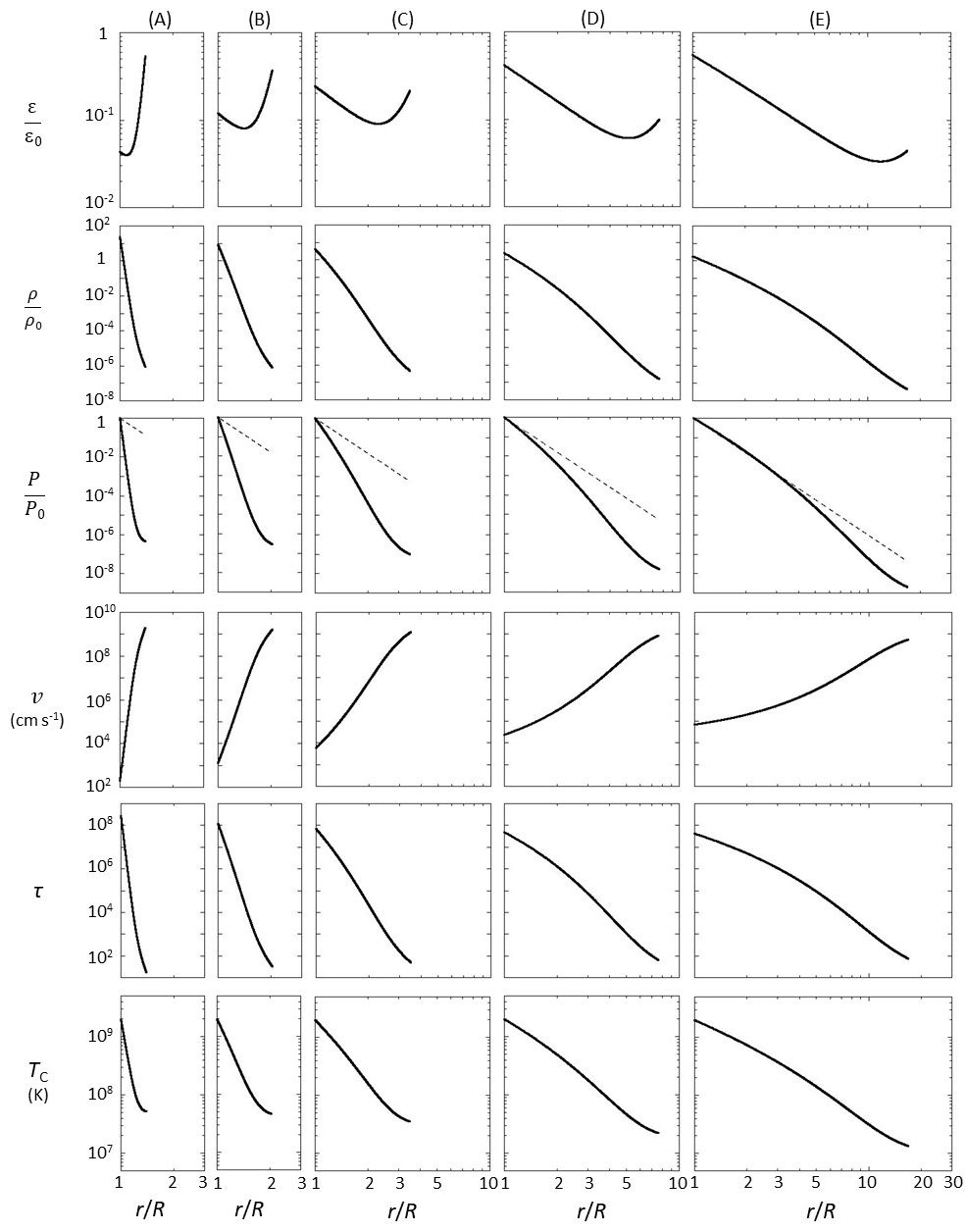} 
\end{center}
\caption{Solved distributions of $\varepsilon$, $\rho$, $P$, $v$, $\tau$ and $T_{\rm C}$ 
as functions of $r$ in five $\dot{M}$ cases for $B_{\rm r, *} = 10^{12}$ G.   Columns (A), (B), (C), (D) and (E) correspond to cases of $\dot{M} = 3 \times 10^{16}$ g s$^{-1}$,
$10^{17}$ g s$^{-1}$, $3 \times 10^{17}$ g s$^{-1}$, $10^{18}$ g s$^{-1}$, and 
$3 \times 10^{18}$ g s$^{-1}$, respectively.  Dashed lines in the row of $P$ represent 
the $r$ dependence of the magnetic pressure as $\propto r^{-6}$. }\label{Solutions_B1}
\end{figure}

\begin{figure}
\begin{center}
\includegraphics[width=15cm]{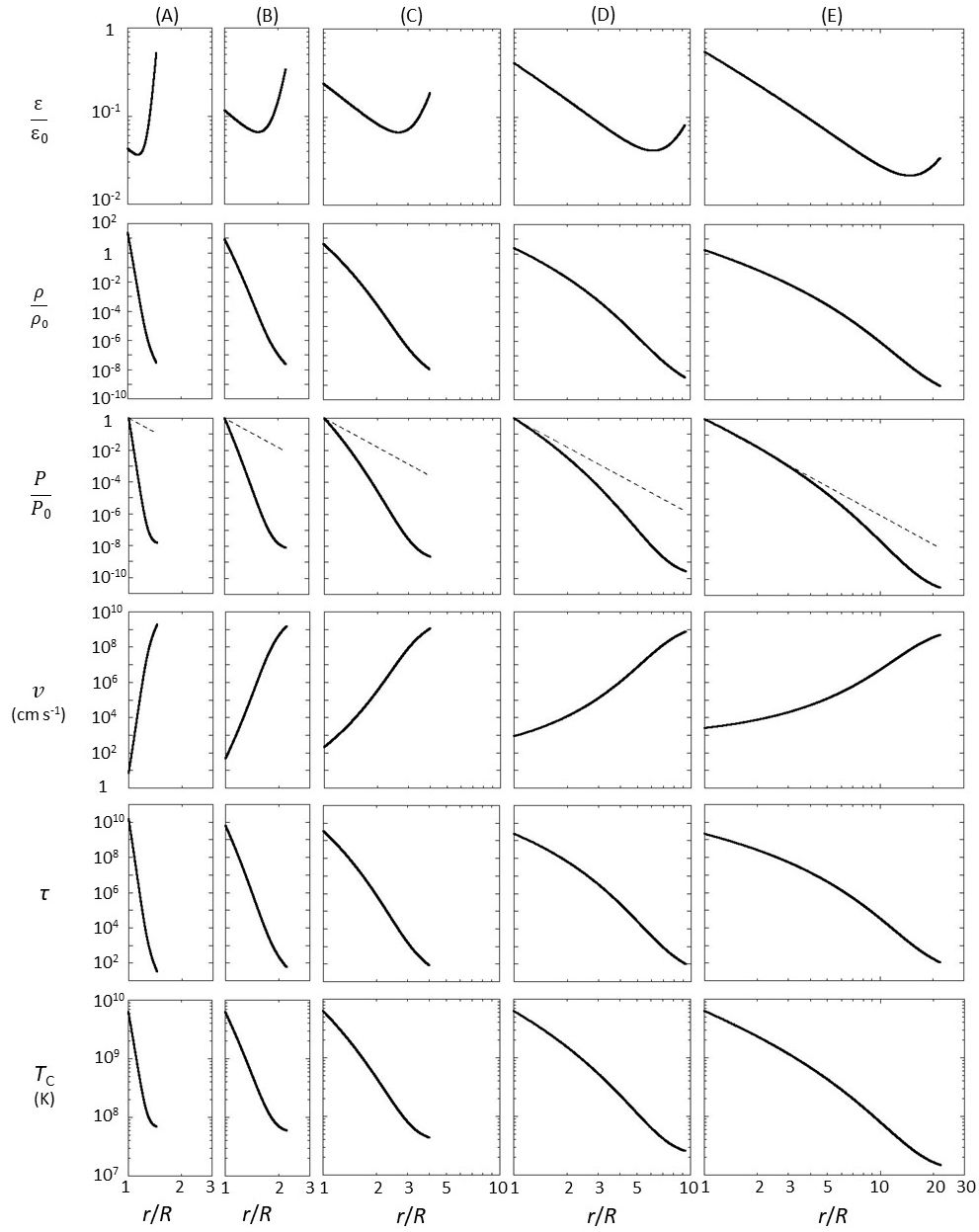} 
\caption{The same as figure \ref{Solutions_B1} but for $B_{\rm r, *} = 10^{13}$ G.}\label{Solutions_B10}
\end{center}
\end{figure}

Solutions of equations (\ref{eqn:DifEq1}) and (\ref{eqn:DifEq2}) have been gotten  
with the procedures described above in various cases of $\dot{M}$,
where we have adopted $M = M_{\odot}$, $R = 10^{6}$ cm and two cases of 
$B_{\rm r, *}$ as 10$^{12}$ G and 10$^{13}$ G.
The obtained $\varepsilon$, $\rho$, $P$, $v$, $\tau$ and $T_{\rm C}$ 
as functions of $r$ are shown 
on five cases of $\dot{M}$ for $B_{\rm r, *} = 10^{12}$ G in figure \ref{Solutions_B1} 
and for $B_{\rm r, *} = 10^{13}$ G in figure \ref{Solutions_B10}, respectively.
Here, $v$ has been calculated 
from the continuity equation.
$\tau$ is the optical depth for the Thomson scattering in the 
tangential direction of the polar cone.
$T_{\rm C}$ is the typical temperature of the matter in the polar cone and has been 
calculated by an equation as $T_{\rm C} = (\rho \; \varepsilon/a)^{1/4}$, where 
$a$ is the first radiation constant.
The normalization parameters in the figures are defined as 
$\varepsilon_{0} = GM/R$, $\rho_{0} = 3 B_{\rm r, *}^{2} R / (8 \pi GM) \sim 9 \times 10^{2} (B_{\rm r, *}/10^{12}$ G)$^{2} (M/M_{\odot})^{-1} (R/10^{6}$  cm) g cm$^{-3}$, 
and $P_{0} = P_{*}$ in equation (\ref{eqn:BtmBC}).

%%%%% figure {Solutions_B1} %%%%%

%%%%% figure {solutions_B10} %%%%%
\begin{figure}
\begin{center}
\includegraphics[width=8cm]{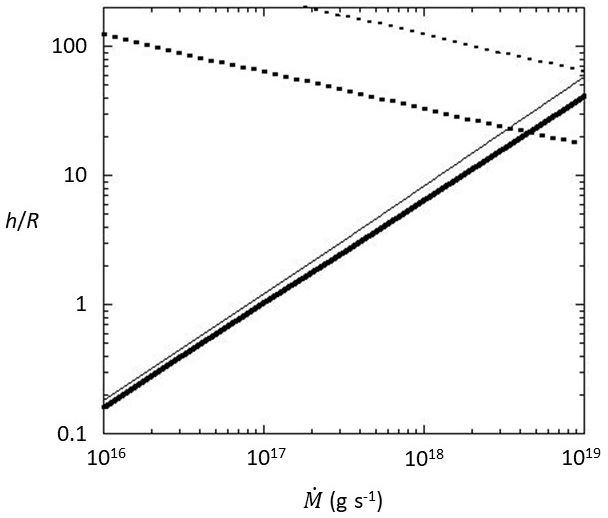} 
\end{center}
\caption{Height of of the polar cone region as a function of $\dot{M}$ (solid lines).  Dotted lines indicate positions of $r_{\rm M,p}$.  The thick and thin lines correspond to cases of $B_{\rm r, *} = 10^{12}$ G and $10^{13}$ G respectively.}\label{Mdot-Height_Relations}
\end{figure}

The height of the polar cone, $h$, defined as 
\begin{equation}
h = r_{\rm S} - R,
\label{eqn:Eq_h}
\end{equation}
is plotted as a function of $\dot{M}$ in figure \ref{Mdot-Height_Relations}.
As seen from this figure, $h$ increases roughly in proportion to $\dot{M}$.
When $\dot{M}$ is $\sim 10^{16}$ g s$^{-1}$, $h$ is about a tenth as small as 
the neutron star radius.
When $\dot{M}$ is $\sim 10^{18}$ g s$^{-1}$, however, 
$h$ gets about ten times as large as the neutron star radius. 

%%%%% figure {Mdot-Height} %%%%%%

%%%%% subsubsection {ELossPC}
\subsubsection{Energy loss from the polar cone}
\label{ELossPC}

From figures \ref{Solutions_B1} and \ref{Solutions_B10}, 
we see, irrespectively of $\dot{M}$, 
that $\varepsilon$ once decreases as $r$ decreases from $r_{\rm S}$ 
but that it turns to increase after $r$ gets smaller than a certain value. 
The variation of $\varepsilon$ is determined by equation (\ref{eqn:DifEq1}) in which 
the first term and the second term in the right hand side are the energy loss rate 
through the radiative diffusion and the energy gain rate from the gravitational energy, 
respectively.
In the upper region near the shock, the radiative cooling rate is dominant to the 
energy gain rate and $\varepsilon$ decreases as the matter flows downwards.
The energy gain rate through the gravitational acceleration, however, becomes 
dominant to the energy loss rate and $\varepsilon$ starts increasing with $r$-decrease 
when $r$ gets smaller than the boundary distance, $r_{\rm B}$, 
where $d\varepsilon/dr = 0$.
Hereafter, the region where $r > r_{\rm B}$ and that where $r < r_{\rm B}$ are called 
as the upper polar cone region and the lower polar cone region respectively.

Let the total energy per mass of the inflow matter in the polar cone be $u$, 
it is expressed as 
\begin{equation}
u = \frac{4}{3} \; \varepsilon - \frac{GM}{r},
\label{eqn:Def_u}
\end{equation}
where the first term in the right side is the enthalpy per mass 
under the situation in which the radiative energy density is largely dominant 
to the gaseous energy density.
Since the top boundary condition of $\varepsilon$ is given as in equation (\ref{eqn:OuterBC}), we have 
\begin{equation}
u_{\rm S} = 0,
\label{eqn:uS}
\end{equation}
just behind the shock front.  This is the result from an approximation 
that the specific total energy, $u$ is just zero at infinity.
Then, if $u$ decreases to $u_{\rm B}$ as the matter falls down to 
the position, $r_{\rm B}$, the inflow matter should lose an amount of energy per mass, 
$w_{\rm U}$ in the upper region as
\begin{equation}
w_{\rm U} = u_{\rm S}-u_{\rm B}=-u_{\rm B}.
\label{eqn:w_U}
\end{equation}
Similarly, another amount of energy per mass, $w_{\rm L}$, 
which is lost from the inflow matter in the lower polar cone region, 
is given as
\begin{equation}
w_{\rm L} = u_{\rm B}-u_{*},
\label{eqn:w_L}
\end{equation}
where $u_{*}$ is the $u$ value at the bottom of the polar cone.
The inflow matter should finally spread over the neutron star surface and 
assimilate with the interior matter of the neutron star.
In order to do that, however, the inflow matter should throw away the radiative thermal energy, $\varepsilon_{*}$, which has remained at the bottom of the polar cone.
This $\varepsilon_{*}$ should be conveyed to the polar mound region as
discussed in the next sub-section and finally be radiated away from its surface.
The amount of the thermal energy, $w_{*}$, which has been carried without radiated away even to the bottom of the polar cone, is given as 
\begin{equation}
w_{*} = \frac{4}{3}\; \varepsilon_{*}.
\label{eqn:w_*}
\end{equation}

\begin{figure}
\begin{center}
\includegraphics[width=8cm]{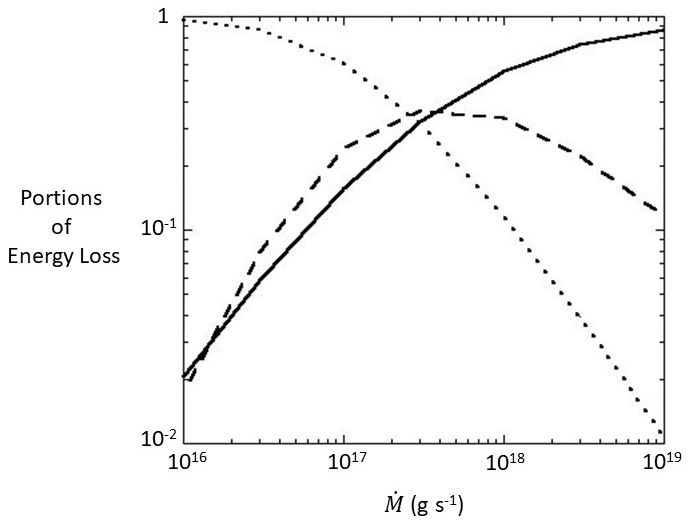} 
\end{center}
\caption{Portions of energy loss in the upper polar cone region (dotted line), the lower polar cone region (dashed line) and in the mound region (thick line) as functions of  $\dot{M}$ in case of $B_{\rm r, *} = 10^{12}$ G.}
\label{EnergyLossPortions}
\end{figure}

Figure \ref{EnergyLossPortions} shows respective portions of the three amounts of the energy, $w_{\rm U}$, 
$w_{\rm L}$ and $w_{*}$ to the sum of the three, as functions of $\dot{M}$ in case of 
$B_{\rm r, *} = 10^{12}$ G.  
The result is about the same in case of $B_{\rm r, *} = 10^{13}$ G. 
We see that $w_{\rm H}$ is dominant when $\dot{M} \lesssim 10^{17}$ g s$^{-1}$, 
the three are comparable to one another when $10^{17}$ g s$^{-1}$ $\lesssim \dot{M} \lesssim 10^{18}$ g s$^{-1}$, and $w_{*}$ is dominant when $\dot{M} \gtrsim 10^{18}$ g s$^{-1}$.

%%%%% figure {EnergyLossPortions} %%%%%

%%%%% subsection {XEPM} %%%%%%%%%%%
\subsection{X-ray emitting polar mound region}\label{XEPM}
\begin{figure}
\begin{center}
\includegraphics[width=8 cm]{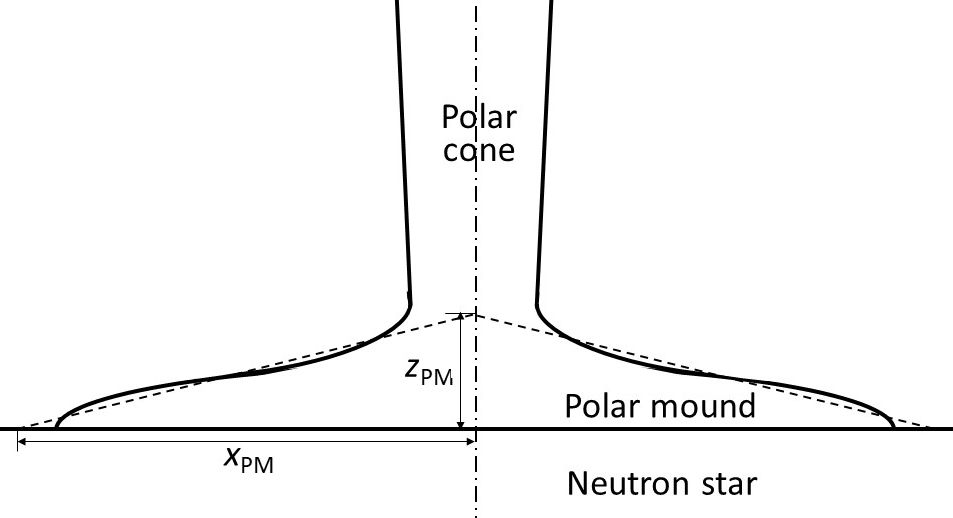} 
\end{center}
\caption{Schmatic cross section of the polar mound.  The shape is approximated by a low cone, as indicated with dashed lines, with a radius, $x_{\rm PM}$ of the base and a height, $z_{\rm PM}$, at the center, in order to roughly estimate the sizes of the polar mound.}
\label{PolarMoundFigure}
\end{figure}

The accreted matter still tends to flow downwards with significant amount of thermal energy 
at the bottom of the polar cone, 
but the thermal pressure should get stronger than the magnetic pressure 
of the surrounding magnetic funnel if the matter flows further 
below the bottom boundary.
Thus, the matter should start expanding along the surface of neutron star, 
dragging the magnetic lines of force.
We call this tangentially expanding region as the polar mound region.
The schematic cross section of the polar mound is drawn in figure \ref{PolarMoundFigure}.

%%%%% figure {PolarMoundFigure} %%%%%

The structure of the polar mound should be determined 
from the following two conditions.

The first is that the matter flowing in the polar mound should radiate away 
the residual thermal energy which has remained at the bottom of the polar cone 
before settling on the surface of the neutron star.
This condition can be represented by a simple equation as
\begin{equation}
\frac{\dot{M}}{2} = \frac{M_{\rm PM}}{t_{\rm D, PM}}.
\label{eqn:EnEq_PM}
\end{equation}
$M_{\rm PM}$ is the total mass in the polar mound, approximately given as
\begin{equation}
M_{\rm PM} = \frac{1}{3} \pi x_{\rm PM}^{2} z_{\rm PM} \rho_{\rm PM},
\label{eqn:M_PM}
\end{equation}
and $t_{\rm D, PM}$ is the photon diffusion time in the direction perpendicular to the 
stellar surface, roughly written as
\begin{equation}
t_{\rm D, PM} = \frac{\kappa_{\rm T} \rho_{\rm PM} z_{\rm PM}^{2}}{3c}.
\label{eqn:t_DPM}
\end{equation}
Here, we have assumed the shape of the polar mound to be a very low cone 
with a radius, $x_{\rm PM}$ of 
the base and a height, $z_{\rm PM}$.
It is simply assumed 
that the density is uniform in the polar mound and that $t_{\rm D, PM}$ is 
represented with the value in the envelope of the cone with the average thickness,
$z_{\rm PM}/3$.
From above three equations we get
\begin{equation}
x_{\rm PM}^{2} = \frac{\dot{M} \kappa_{\rm T}}{2\pi c} z_{\rm PM}.
\label{eqn:x-z_Rel_1}
\end{equation}

The second condition is a pressure balance between the radiation pressure 
pushing out the magnetic lines of force and 
the magnetic pressure intensified by being dragged by the expanding flow in the polar mound. 
The radiation pressure, $P$, could be approximately obtained by extending 
the $P$ - $r$ relation in the polar cone 
to a position slightly lower than the bottom of the polar cone.
If we express it as $P \propto r^{-\alpha}$, $P$ which pushes out the magnetic lines 
of force is roughly given as
\begin{equation}
P = P_{*} \left( \frac{R}{R + z_{\rm PM}} \right)^{-\alpha},
\label{eqn:P_PM}
\end{equation}
where $P_{*}$ is the pressure at the bottom of the polar cone as is given in equation 
(\ref{eqn:BtmBC}).
The magnetic pressure which pushes back the matter pressure could be approximated  as
\begin{equation}
\left( \frac{B^{2}}{8\pi} \right)_{\theta} = \left( \frac{B^{2}}{8\pi} \right) \frac{z_{\rm PM}}{(x_{\rm PM}^{2} + z_{\rm PM}^{2})^{1/2}} = P_{*} \frac{(x_{\rm PM}^{2} + z_{\rm PM}^{2})^{1/2}}{z_{\rm PM}},
\label{eqn:MagP_PM}
\end{equation}
where we have assumed
\begin{equation}
B = B_{\rm r, *} \frac{(x_{\rm PM}^{2} + z_{\rm PM}^{2})^{1/2}}{z_{\rm PM}}.
\label{eqn:B-Br*}
\end{equation}
By equating two equations (\ref{eqn:P_PM}) and (\ref{eqn:MagP_PM}), we get
\begin{equation}
\left( \frac{R}{R + z_{\rm PM}} \right)^{-\alpha} = \frac{(x_{\rm PM}^{2} + z_{\rm PM}^{2})^{1/2}}{z_{\rm PM}}.
\label{eqn:x-z_Rel_2}
\end{equation}

\begin{figure}
\begin{center}
\includegraphics[width=8cm]{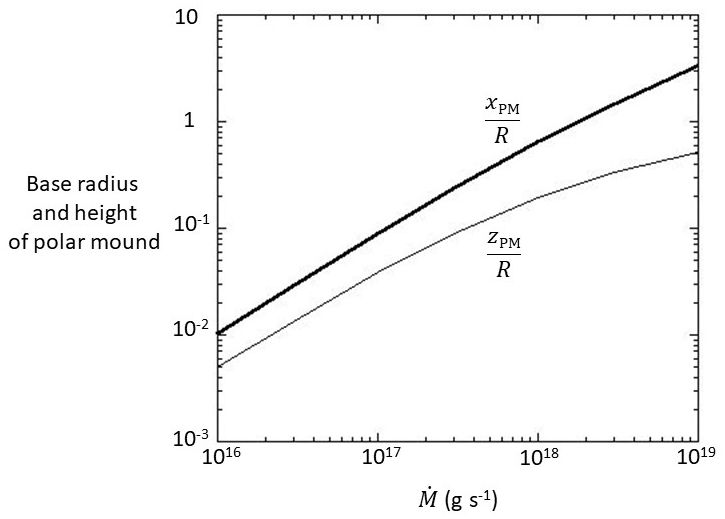} 
\caption{Base radius (the thick line)  and height (the thin line) of the polar mound, in unit of the stellar radius, as a function of  $\dot{M}$.  }
\label{MoundParameters}
\end{center}
\end{figure}

By seeing $\alpha$ from the $P$ distribution in the polar cone 
as a function of $\dot{M}$, shown in figures \ref{Solutions_B1} and \ref{Solutions_B10},  
and  solving simultaneous equations (\ref{eqn:x-z_Rel_1}) and (\ref{eqn:x-z_Rel_2}), 
we obtain $x_{\rm PM}$ and $z_{\rm PM}$ as functions of $\dot{M}$ in two cases of $B_{*}$.
The results in case of $B_{*} = 10^{12}$ gauss are plotted  in figure \ref{MoundParameters}.  Those in case of $B_{*} = 10^{13}$ gauss are almost identical.
When $\dot{M}$ becomes larger than $10^{18}$ g s$^{-1}$, $x_{\rm PM}$ comes to 
be as large as or even larger than $R$ and simple arguments as done above become inapplicable.

%%%%% figure {MoundParameters} %%%%%

%%%%% section {X-ray_Emissions} %%%%%%%%%%%%%%%%%%%%
\section{X-ray emissions from the magnetic polar regions}\label{X-ray_Emissions}

We have seen the portions of energy loss from the upper polar cone region, the lower polar cone region 
and the polar mound region in \ref{ELossPC}.
Since the energy lost from these regions should be emitted mainly in X-rays 
from their surfaces, 
the portions of the energy loss from the three 
regions as shown in figure \ref{EnergyLossPortions} correspond to those of the local 
X-ray luminosities.

In the polar cone region, photons are considered to diffuse out from the side surface of 
the cone.
The differential X-ray luminosity emitted from the polar cone surface 
per unit radial-length is given from equation (\ref{eqn:BernoulliEq}) as 
\begin{equation}
\frac{dL}{dr} = \frac{4\pi c}{3 \kappa_{\rm T}} \varepsilon.
\label{eqn:dLdr}
\end{equation}
It should be noted that this quantity is proportional to $\varepsilon$ 
and does not directly depend on $\dot{M}$ nor $\Theta$.

Since thermal equilibrium between matter and radiation is considered to be well 
established in the polar cone, emission from the surface of the polar cone should be basically blackbody emission.
In that case, the effective temperature, $T_{\rm e}$, of a surface of the polar cone 
at a distance, $r$, is estimated from the following equation as
\begin{equation}
\frac{dL}{dr} = 2 \pi r \Theta \sigma T_{\rm e}^{4},
\label{eqn:RltnTe^4}
\end{equation}
where $\sigma$ is the Stephan Boltzmann constant.

However, we should take account of the dilution effect of the blackbody photons, 
since scattering is considered to be dominant to absorption in such an atmosphere 
with a temperature as high as 10$^{7}$ K as the surface of the polar cone.
In such a situation, we can approximate that the thermal equilibrium between matter 
and radiation is established within a region with an optical depth for the scattering, 
$\tau_{\rm a} \gg 1$, and only elastic scatterings take place in the atmosphere with 
$\tau < \tau_{\rm a}$.
Under this approximation, the observed spectrum can be approximated with 
the blackbody spectrum with the apparent temperature, $T_{\rm a}$, 
at the bottom of the scattering atmosphere with $\tau_{\rm a}$.
In the atmosphere with $\tau \lesssim \tau_{\rm a}$, a radiation energy flux should 
be conserved and we can approximate that
\begin{equation}
\frac{T_{\rm a}^{4}}{\tau_{\rm a}} = T_{\rm e}^{4}.
\label{eqn:Rltn_Ta&Te}
\end{equation} 

From equations (\ref{eqn:dLdr}) and (\ref{eqn:RltnTe^4}) with help of equation (\ref{eqn:Rltn_Ta&Te}), the observed blackbody (apparent) temperature, $T_{\rm a}$, 
is written as
\begin{equation}
T_{\rm a} = \left( \frac{2c}{3\kappa_{\rm T} \sigma} \frac{\tau_{\rm a} \varepsilon}{r \Theta} \right)^{1/4}.
\label{eqn:Ta}
\end{equation}

If we introduce the following relations in general as
\begin{eqnarray}
T_{\rm a} &\propto& r^{-p}, \label{eqn:T-r} \\
\varepsilon &\propto& r^{\nu}, \label{eqn:epsilon-r} \\
\Theta &\propto& r^{\eta}, \label{eqn:theta-r} \\
\tau_{\rm a} &\propto& T_{\rm a}^{\omega}, \label{eqn:tau-Tc} 
\end{eqnarray}
we get
\begin{equation}
p = \frac{-\nu + \eta + 1}{4 - \omega}.
\label{eqn:Rltn_p}
\end{equation}
Note that differential equations  (\ref{eqn:DifEq1}) and (\ref{eqn:DifEq2}) in  
\ref{Structure} does not include the parameter $\Theta$.  
Thus, the solutions as shown in figures \ref{Solutions_B1} and \ref{Solutions_B10} 
allow such a freedom of $r$ 
dependence of $\Theta$ as in equation (\ref{eqn:theta-r}), as far as $\Theta \ll 1$.
In this study, we have assumed $\eta = 0.5$ as in equation (\ref{eqn:Theta_r/R}).

As seen in top panels of figures \ref{Solutions_B1} and \ref{Solutions_B10}, 
two local maximums of $\varepsilon$
appear, and the locally most luminous positions in the polar cone are 
the top of the polar cone just behind the shock and the bottom of the polar cone just on the neutron star surface. 
The most luminous place, however, depends on $\dot{M}$.
When $\dot{M} \lesssim  10^{17}$ g s$^{-1}$, the top of the polar cone is the most luminous, 
when $\dot{M} \simeq 3 \times 10^{17}$ g s$^{-1}$, the top and the bottom of the polar cone are 
comparably luminous, and 
when $\dot{M} \gtrsim  10^{18}$ g s$^{-1}$, the bottom is the most luminous.  

The relative amount of the differential luminosity between the top and bottom of the polar cone  
is seen to depend on $\dot{M}$ in figure \ref{EnergyLossPortions} too.
This figure further reminds us that the polar mound region is the most luminous  
when $\dot{M} \gtrsim 3 \times 10^{17}$ g s$^{-1}$.

The X-ray luminosity from the polar mound region, $L_{\rm PM}$, is given as
\begin{equation}
L_{\rm PM} = \frac{\dot{M}}{2} w_{*},
\label{eqn]L_PM}
\end{equation}
where $w_{*}$ has been defined in \ref{ELossPC}.
In the polar mound region, the matter should successively sink into the surface layer
of the neutron star as it expands tangentially on the stellar surface.
Because of the matter sinking and the area-extension,
the column density should decreases and it get easier for photons 
to diffuse out as the distance from the mound center increases.
Furthermore, since the magnetic lines of force are dragged tangentially 
along the surface of the polar mound, photons easily go outward, without 
scattered by ambient electrons which cannot move across the dragged lines of force.
As a result, we can expect that the X-ray emission could be largely weighted on the 
outermost area of the polar mound and be approximated by a single blackbody emission 
with an apparent temperature, $T_{\rm a, PM}$, which could be calculated from the following 
equation as
\begin{equation}
L_{\rm PM} = \chi \pi x_{\rm PM}^{2} \sigma T_{\rm a, PM}^{4}.
\label{eqn:T_aPM^4}
\end{equation}
Here, $\chi$ is a parameter which represents both a reducing factor of the effective 
emission area from the whole surface area of the mound 
and a dilution factor of the scattering dominant atmosphere, and should be $\ll 1$.

%%%%% subsection {X-ray_spectrum} %%%%%
\subsection{Properties of X-ray spectrum}\label{X-ray_spectrum}
When $\dot{M} \lesssim 10^{17}$ g s$^{-1}$, X-ray emission from the upper polar cone is 
the most luminous.
As seen from the top panels of figures \ref{Solutions_B1} and \ref{Solutions_B10}, the logarithmic slope of $\varepsilon$ against $r$, $\nu$, is much larger than unity, and thus a single blackbody 
spectrum is expected in this $\dot{M}$ range within the simplified arguments in this study.
Since establishment of the radiation field is thought to be insufficient behind the shock 
as in figure \ref{Mdot-zeta}, however, 
Comptonization of bremsstrahlung photons and blackbody photons produced around there should be taken into account to consider properties 
of X-ray spectrum from there, which was already studied by Becker and Wolff (2007) 
and Wolff et al. (2016).
We do not go into detail on this low $\dot{M}$ range here.

When $\dot{M} \gg 10^{17}$ g s$^{-1}$, on the other hand, the X-ray emission 
from the upper polar cone region get less luminous than 
those from the lower polar cone  region 
and the polar mound region.
Furthermore, since the distance of the shock front, $r_{\rm S}$, is several times 
as large as the neutron star radius in case of $\dot{M} \simeq 10^{18}$ g s$^{-1}$ 
(see figure \ref{Mdot-Height_Relations}), the temperature of the emission from the upper polar cone region 
should be much lower than those from the two regions near the neutron star surface.
Thus, we neglect contribution of the emission from the upper polar cone region here 
because our present concern exists on properties of X-ray spectrum in the energy range above a few keV in the $\dot{M}$ range around 10$^{18}$ g s$^{-1}$.

From the lower polar cone region, we can expect a multi-color blackbody spectrum,
which should be expressed with the $p$-free disk model (Kubota \& Makishima 2004).
This model was developed to reproduce X-ray spectra from accretion disks and 
the surface blackbody temperature is assumed to distribute as a function of 
an orbital radius in a disk as in equation (\ref{eqn:T-r}).
In the present case considered here, the $p$ parameter is given by equation 
(\ref{eqn:Rltn_p}).
The value of $\nu$, which is defined as the logarithmic slope of $\varepsilon$ against $r$ 
in equation (\ref{eqn:epsilon-r}), can be calculated from the radial distribution of $\varepsilon$ 
solved in \ref{Structure}, and is $\sim$ -1.3 near the bottom of the polar cone 
when $\dot{M}$ is around 10$^{18}$ g s$^{-1}$.
Then,  $p$ is estimated to be $\sim$ 0.7, taking account of $\eta$ = 0.5 as assumed 
in equation (\ref{eqn:Theta_r/R}) and setting $\omega$ = 0 for simplicity.
The highest apparent temperature in the multi-color component can be estimated 
from equation (\ref{eqn:Ta}) by setting $\varepsilon = \varepsilon_{*}$, $r = R$ and
$\Theta = \Theta_{*}$, and assuming $\tau_{\rm a}^{1/4} \simeq 2$ (e.g. Shimura \& Takahara 1995).  Since $\varepsilon_{*} \simeq 0.4 \varepsilon_{0}$ 
when $\dot{M} = 10^{18}$ g s$^{-1}$ as seen both in figures \ref{Solutions_B1} and \ref{Solutions_B10}, we get the highest apparent temperature 
$\sim$ 5 keV around that accretion rate.

From the polar mound region, a quasi-single blackbody spectrum should be detected. 
The apparent temperature, $T_{\rm a,PM}$, is calculated from equation (\ref{eqn:T_aPM^4}) to be again $\sim$ 5 keV when $L_{\rm PM} \simeq 10^{38}$ 
erg s$^{-1}$ and $\chi^{-1/4} (x_{\rm PM}/R)^{-1/2} \simeq 2$. 

According to above arguments, we can expect a hybrid spectrum composing of 
the multi-color blackbody spectrum from the lower polar cone region 
and the quasi-single blackbody spectrum from the polar mound region, 
for X-ray emissions in the energy range above a few keV, 
in case of $\dot{M} \gg 10^{17}$ g s$^{-1}$. 

Observationally, Makishima et al. (1999) showed that continuum X-ray spectra of 
X-ray pulsars in 2 - 50 keV can generally be reproduced with the NPEX model, 
in which the model spectrum, $F(E)$, is assumed as
\begin{equation}
F(E) = ( G_{1} E^{-\lambda} + G_{2} E^{2} ) \exp \left( - \frac{E}{kT} \right),
\label{eqn:NPEX}
\end{equation}
where $E$ is a photon energy and $G_{1}$ and $G_{2}$ are normalization factors.
In the present framework, the first and second components could be understood to be 
the multi-color blackbody spectrum from the bottom polar cone region 
and the quasi-single blackbody spectrum from the polar mound region
respectively.
The multi-color blackbody spectrum from the polar cone 
has a power-law like spectrum in the medium energy range 
and an exponential decay determined by the highest surface temperature of the polar cone in the high energy range.
This could well be reproduced by the first component of the NPEX model.
The second component of the NPEX model has the Rayleigh-Jeans spectrum in the low 
energy range and the Wien spectrum in the high energy range, and could be considered 
to represent the quasi-single blackbody spectrum from the polar mound surface.
Although there is no reason why the first and second components have the common 
temperature in these considerations,  the NPEX model is the phenomenological model to reproduce the observed spectra and it would not be required to separate the common temperature into two in practical spectral fits. 

From these considerations, the fact that the NPEX model consisting of the two spectral components, the cut-off power law spectrum  and the single blackbody-like spectrum, 
can well reproduce the continuum spectra observed from many X-ray pulsars 
could be considered to strongly support the present model which predicts presences 
of the two spectral components, the multi-color blackbody spectrum from the polar 
cone region and the quasi-single blackbody spectrum from the polar mound region.
It should be noted that the temperatures obtained with the NPEX model fits to X-ray 
spectra observed from several sources by Makishima et al. (1999) distribute around 5 keV, which agree to the values estimated above in the present study.

Furthermore, Kondo, Dotani \& Inoue (in preparation) have recently analyzed 
phase-resolved spectra observed from Her X-1 with Suzaku 
and have successfully resolved the continuum spectra into 
three components, each of   
which keeps its respective spectral shape constant 
and varies only its normalization factor in association with the pulse-phase change, 
independently of the other components. 
A remarkable thing is that the two components resolved in the energy range 
above $\sim$ 2 keV are well reproduced with the multi-color blackbody spectrum 
and the single blackbody spectrum respectively, although they have been obtained 
in a model-independent way.

%%%%% subsection {X-ray_pulse_profiles} %%%%%%%%%%
\subsection{Properties of X-ray pulse profile}\label{X-ray_pulse_profiles}
Here, we discuss pulse profiles expected from the presently considered configurations 
in the magnetic polar regions in case of $\dot{M} \gg 10^{17}$ g s$^{-1}$.

As seen in figure 12-(c) in Makishima et al. (1999), the crossover energy, 
where the two components in the NPEX model in equation (\ref{eqn:NPEX}) cross over, generally exists around 10 keV.
Thus, we can consider that the multi-color blackbody component, identical to the 
first component in the NPEX model, from the lower polar cone should mainly govern 
the pulse profile in an energy range, say, 2 to 6 or 7 keV.
On the other hand, in an energy range above say 15 or 20 keV, the single blackbody 
component, identical to the second component in the NPEX model, from the polar 
mound region should be the main component in the pulse profile.

We discuss pulse profiles expected for X-rays from the polar cone and the polar mound respectively, below.
In calculating the pulse profiles, we take into account two effects deforming the profiles, 
occultation of the emission regions by the neutron star body and 
influences of relativistic light bending.
They are explained in appendix.

\subsubsection{Directional distributions of X-ray emissions from the polar cone}
X-rays from the polar cone mainly forms the pulse profile in 2 - 6 $\sim$ 7 keV.

Several X-ray pulsars exhibit absorption features at Cyclotron resonance energies 
around 20 $\sim$ 40 keV in their X-ray spectra (e.g. Makishima et al. 1999).
Since X-ray energies of 2 to 6 $\sim$ 7 keV considered now are significantly lower 
than these Cyclotron energies, X-rays in this energy range cannot move electrons 
in the direction perpendicular to the magnetic lines of force sufficiently.
Hence, when X-rays diffuse out in the azimuthal direction of the polar cone, 
those having a linear polarization perpendicular to the magnetic lines of force 
(``ordinary photons'') can easily 
go out of the surface layer without suffering from electron scattering, but those 
having another polarization in direction to the line of forces (``extraordinary photons'') 
cannot escape without a number of electron scatterings from the surface layer.
Even the extraordinary photons, however, become free from electron scattering, 
if they are scattered in the direction to the magnetic lines of force. 
Therefore, the extraordinary photons should gradually be focused in the 
direction to the magnetic lines of force through a number of electron scatterings.
This beaming effect was first pointed out by Basko and Sunyaev (1975).

As a result, photons diffusing out in the azimuthal direction from the central region of 
the polar cone
could be divided into two groups in the surface layer:
One is the ordinary photon group and could be emitted with a constant intensity 
per a solid angle to every direction from the surface.
Then, a differential flux per unit length of the polar cone, 
$dF_{\rm O}/dr$, detected by an observer 
in a direction with an angle, $\phi$, from the outward direction of the central axis 
of the polar cone at a distance of $D$ can be expressed as
\begin{equation}
\frac{dF_{\rm O}}{dr} = \frac{4}{\pi} \frac{dL_{\rm PC, O}/dr}{4\pi D^{2}} \; \sin \phi,
\label{eqn:Def_F_O}
\end{equation}
where $dL_{\rm O}/dr$ is a differential luminosity of this ordinary component 
emitted from the surface per unit length.

The other half is the extraordinary photon group and should be focused in the direction 
of the surface magnetic lines of force.
If the beam pattern is assumed to have a form as $\exp [-(\phi/\sigma_{\rm E})^{2}]$
with a parameter,  $\sigma_{\rm E}$, representing a beam width, 
the differential flux per unit length of the polar cone, 
$dF_{\rm E}/dr$ can be approximated 
under a condition $\sigma_{\rm E} \ll 1$ as 
\begin{equation}
\frac{dF_{\rm E}}{dr} = \frac{dF_{\rm E, 0}}{dr}  \; exp[-(\frac{\phi}{\sigma_{\rm E}})^{2}] \; \sin \phi ,
\label{eqn_F_E}
\end{equation}
where
\begin{equation}
\frac{dF_{\rm E, 0}}{dr} = \frac{4}{\pi^{1/2} \sigma_{\rm E}^{3}} \frac{dL_{\rm PC, E}/dr}{4 \pi D^{2}}.
\label{eqn_F_E0}
\end{equation}
$dL_{\rm PC, E}/dr$ is the differential luminosity of this extraordinary component.
Calculated from equation (\ref{eqn_F_E}), $dF_{\rm E}/dE$ has a peak at 
$\phi_{\rm P}$ given as 
\begin{equation}
\phi_{\rm P} = \frac{\sigma_{\rm E}}{\pi^{1/2}}.
\label{eqn:phi_P}
\end{equation}

Let us introduce  $dL_{\rm PC}/dr$ as the total differential luminosity per unit length of the polar cone 
and set $dL_{\rm PC, O}/dr = dL_{\rm PC, E}/dr = (dL_{\rm PC}/dr)/2$, then we get
the total differential flux $dF_{\rm PC}/dr$ as
\begin{equation}
\frac{dF_{\rm PC}}{dr} = \frac{dL_{\rm PC}/dr}{4 \pi D^{2}} \left( \frac{2}{\pi^{1/2} \sigma_{\rm E}^{3}} \; exp[-(\frac{\phi}{\sigma_{\rm E}})^{2}] \; + \; \frac{2}{\pi} \; \right) \; \sin \phi.
\label{eqn:dF_PC}
\end{equation}

\subsubsection{Directional distributions of X-ray emissions from the polar mound}
X-rays from the polar mound mainly forms a pulse profile in the energy range above $\sim$ 20 keV.

In the polar mound, 
the accreted matter produces a mound-like structure by dragging the magnetic 
lines of force and finally lands on the stellar surface there.
The remaining thermal energy at the bottom of the polar cone 
should be carried in the polar mound as radiation energy 
and the photons should diffuse out in the direction to the surface of the mound.  
The X-ray emission from the polar mound could be expected to be largely 
weighted on the outermost area as discussed earlier in this section.
Since magnetic lines of force cover this mound region, the same thing happens to 
the diffusing-out-photons in the surface layer as in the polar cone.
Namely, ordinary photons can easily go out in every directions 
without suffering from electron scattering, 
while extraordinary photons tend to escape, after a number of electron scatterings, 
in the direction along the magnetic lines of force.
In the case of the polar mound, however, the surface magnetic lines of force should 
gradually curve to be normal to the stellar surface 
towards the outermost side of the polar 
mound (see figure \ref{PolarMoundFigure}) 
and the extraordinary photons should tend to go out in directions 
around the normal axis to the stellar surface.
The ordinal photons should, on the other hand, be radiated away, 
independently of the directions of the magnetic lines of force,  
around the normal axis to the stellar surface simply by the projection effect.
We approximate, therefore, that an intensity per solid angle 
from the polar mound surface is isotropic and that the observed flux, $F_{\rm PM}$, 
is written as 
\begin{equation}
F_{\rm PM} = 4 \frac{L_{\rm PM}}{4 \pi D^{2}} \cos \phi,
\label{eqn:F_PM}
\end{equation}
where $L_{\rm PM}$ is the X-ray luminosity radiated from the polar mound and
$\phi$ is an angle between the line of sight and the central axis of the mound. 

\subsubsection{Expected pulse profiles}\label{Expected_Pulse_Profiles}

\begin{figure}
\begin{center}
\includegraphics[width=8cm]{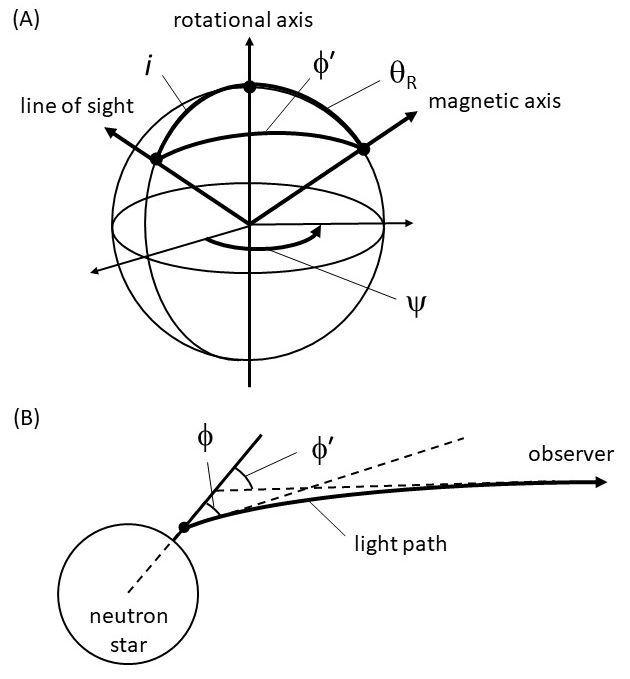} 
\end{center}
\caption{ (A) Angular relations among the rotational axis,  
the magnetic axis and the line of sight, and (B) a relation between $\phi$ and $\phi$' 
reflecting the light bending effect.  }
\label{AngularGeometries}
\end{figure}

A pulse profile of X-rays from the polar cone and the polar mound 
can be obtained from equation (\ref{eqn:dF_PC}) and equation (\ref{eqn:F_PM}) respectively 
by substituting $\phi$ derived from $\phi$', which periodically varies in association with the 
neutron star rotation, taking account of the light bending effect as shown in appendix.
$\phi$' is calculated with an inclination angle, $i$, 
of the line of sight to the rotational axis, 
an angle, $\theta_{\rm R}$, of the magnetic axis of the neutron star 
to the rotational axis, and a rotational angle, $\psi$, of the magnetic axis 
around the rotational axis.
Figure \ref{AngularGeometries} indicates (A) angular relations among the rotational axis,  
the magnetic axis and the line of sight, and (B) a relation between $\phi$ and $\phi$' 
reflecting the light bending effect.  
Note that the opening angle of the polar cone and the height of the polar mound are 
both assumed to be negligibly small in the pulse-profile computations 
and that all the angular distributions of 
the emissions from them are given as functions of $\phi$ referring to the magnetic axis.

%%%%% figure {AngularGeometries} %%%%%

\begin{figure}
\begin{center}
\includegraphics[width=16cm]{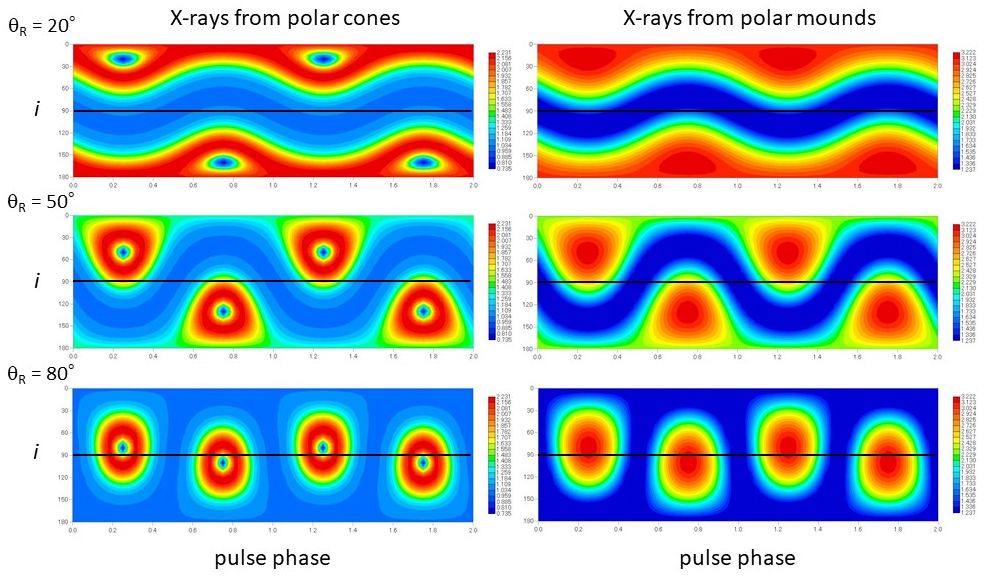} 
\end{center}
\caption{Two dimensional distributions of flux from the polar cone (left) and the polar mound (right) on pulse phases and inclinations in three cases of $\theta_{\rm R}$.  
$\xi$ = 5, $\theta_{\rm R} = 30^{\circ}$, and $\phi_{\rm C} = 60^{\circ}$ are assumed.}
\label{PulseProfileMap}
\end{figure}

\begin{figure}
\begin{center}
\includegraphics[width=15cm]{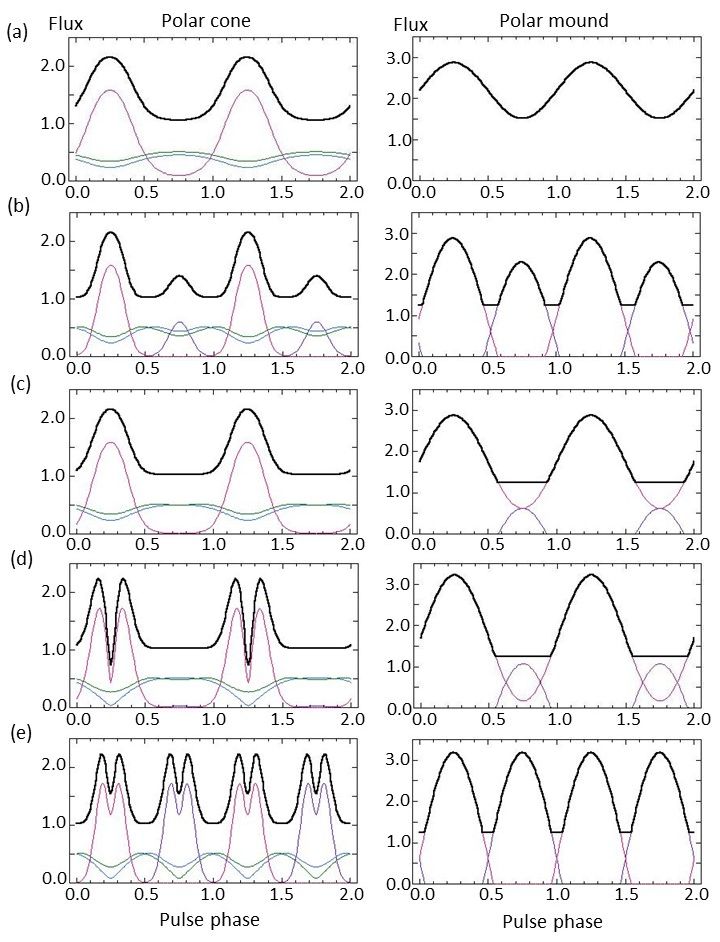} 
\end{center}
\caption{Pulse profiles of fluxes from the polar cone (left) and the polar mound (right) in five cases for the combination of  $i$ and $\theta_{\rm R}$. 
(a) through (e) corresponds to 
cases of (50$^{\circ}$, 20$^{\circ}$), (80$^{\circ}$, 50$^{\circ}$), (60$^{\circ}$, 30$^{\circ}$), (50$^{\circ}$, 50$^{\circ}$) and (90$^{\circ}$, 80$^{\circ}$), respectively. If we replace the values of $i$ and $\theta_{\rm R}$ with each other, the profile is the same.   $\xi = 5$, $\sigma_{\rm E} = 30^{\circ}$ and $\phi_{\rm C} = 60^{\circ}$ are assumed.  The red, purple, blue and green lines in the polar cone component are profiles 
of the pencil beam component from the N-pole, that from the S-pole, the fan-beam component from the N-pole and that from the S-pole, respectively.  The red and purple lines in the polar mound component are the profiles from the N-pole and S-pole, respectively.}
\label{PulseProfileFigures}
\end{figure}

\begin{figure}
\begin{center}
\includegraphics[width=15cm]{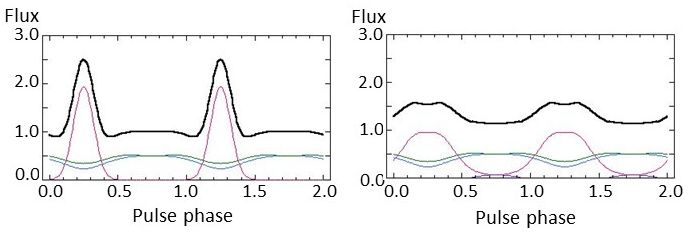} 
\end{center}
\caption{Examples of pulse profile variations in terms of $\sigma_{\rm E}$.
The pulse profile of the polar cone component in figure \ref{PulseProfileFigures}-(c) 
in which $\sigma_{\rm E} = 30^{\circ}$ changes to the left and right profiles in this figure by replacing $\sigma_{\rm E}$ to 20$^{\circ}$ and 40$^{\circ}$ respectively.}
\label{PLS_PRFL_ParameterDependence_1}
\end{figure}

\begin{figure}
\begin{center}
\includegraphics[width=15cm]{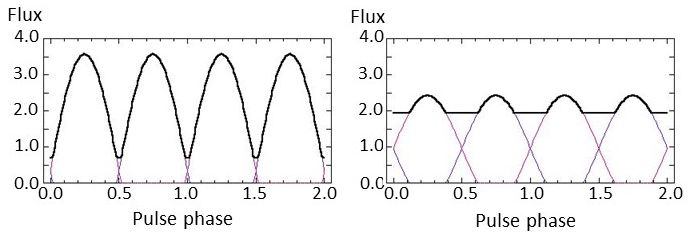} 
\end{center}
\caption{Examples of pulse profile variations in terms of $\xi$.
The pulse profile of the polar cone component in figure \ref{PulseProfileFigures}-(e) 
in which $\xi = 5$ changes to the left and right profiles in this figure by replacing $\xi$ to 10 and 2.5 respectively.}
\label{PLS_PRFL_ParameterDependence_2}
\end{figure}

\begin{figure}
\begin{center}
\includegraphics[width=15cm]{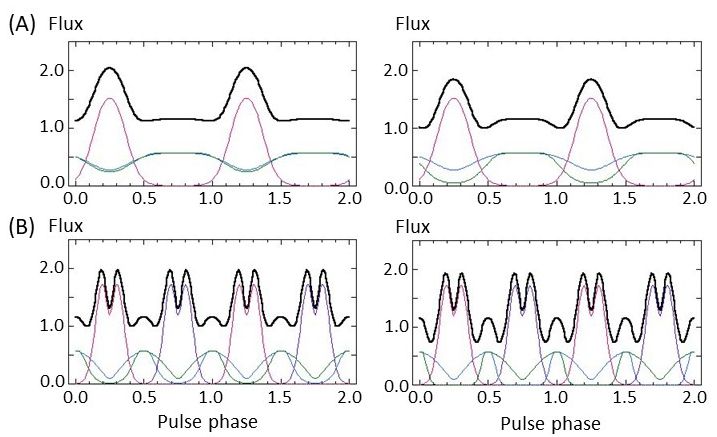} 
\end{center}
\caption{Two examples of pulse profile variations in terms of a combination of $\xi_{\rm S}$ and $\phi_{\rm C}$.
(A) The pulse profile of the polar cone component in figure \ref{PulseProfileFigures}-(c) 
in which $\xi$ = 5 and $\phi_{\rm C} = 60^{\circ}$ changes to the left profiles by replacing $\xi$ to 10, and further changes to the right profile by replacing $\phi_{\rm C}$ to 30$^{\circ}$.
(B) The pulse profile of the polar cone component in figure \ref{PulseProfileFigures}-(e) 
in which $\xi$ = 5 and $\phi_{\rm C} = 60^{\circ}$ changes to the left and right profiles by replacing ($\xi$, $\phi_{\rm C}$) to (10, 30$^{\circ}$) and further to (10, 15$^{\circ}$), respectively.}
\label{PLS_PRFL_ParameterDependence_3}
\end{figure}

Figure \ref{PulseProfileMap} show two dimensional distributions of the flux 
from the polar cone and the polar mound 
respectively as functions of $i$ and $\psi$, in three cases of $\theta_{\rm R}$.
Fluxes in figures \ref{PulseProfileMap} through \ref{PLS_PRFL_ParameterDependence_3} 
are given as relative fluxes to those in cases of isotropic emission, by normalizing  
$dF_{\rm PC}/dr$ in equation (\ref{eqn:dF_PC}) by  $(dL_{\rm PC}/dr)/(4 \pi D^{2})$ for the polar cone component, 
and $F_{\rm PM}$ in equation (\ref{eqn:F_PM}) by $L_{\rm PM}/(4 \pi D^{2})$ 
for the polar mound component, respectively.

Since both emissions from the polar cone and the polar mound 
are mainly focused in directions to 
the two magnetic axes corresponding to the N- and S-poles, two enhancements of 
the flux appear in each of these maps.
In case of emissions from the polar mound, 
each enhancement has a flux-peak at a position 
where the line of sight just directs to the N or S pole.
In case of those from the polar cone, however, 
each enhancement has a crator-like hole at its 
center surrounded by a circular wall with the angular radius of $\phi_{\rm P}$ 
as defined in equation (\ref{eqn:phi_P}).

%%%%% figure {PulseProfileMap} %%%%%

Hereafter, we discuss properties of the pulse profiles only in the range of $i \leq 90^{\circ}$ and let the magnetic axis closer to the line of sight be the N-pole.

In case of those from the polar mound, pulse profiles in terms of $i$ 
is somewhat simpler than that of the polar cone.
When $i$ is larger than a boundary value, two peaks are exhibited, a stronger peak 
from the N-pole side and a weaker peak from the S-pole side. 
When $i$ is smaller than the boundary value, on the other hand, 
we see only one peak from the N-pole side.  
The boundary $i$-value depends on $\theta_{\rm R}$ and gets larger as $\theta_{\rm R}$ gets smaller.

The similar trend of the overall flux-enhancements is seen in pulse profiles of X-rays from the polar cone.
Sub-peaks, however, appear in some ranges of $i$ on these cases.

Let us introduce, here, $\theta_{\rm L, N}$ and $\theta_{\rm L, S}$ as
\begin{equation}
\theta_{\rm L, N} = |\theta_{\rm R}-i|
\label{eqn:theta_LN}
\end{equation}
and 
\begin{equation}
\theta_{\rm L, S} = |\pi - \theta_{\rm R}-i|.
\label{eqn:theta_LS}
\end{equation}
Then, we observe two sub-peaks around the top of the flux enhancement 
associated with the N-pole, 
in case of 
\begin{equation}
\theta_{\rm L,N} < \phi_{\rm P},
\label{eqn:thetaLN<phiP}
\end{equation}
while another couple of sub-peaks appears in the flux enhancement 
associated with the S-pole, 
in case of 
\begin{equation}
\theta_{\rm L,S} < \phi_{\rm P},
\label{eqn:thetaLS<phiP}
\end{equation}
From above four equations (\ref{eqn:theta_LN}) $\sim$ (\ref{eqn:thetaLS<phiP}), 
we see that both enhancements with N- and S- poles exhibit respective 
two sub-peaks only when both $i$ and $\theta_{\rm R}$ are close to $\pi/2$ 
satisfying a condition $i + \theta_{\rm R} < \pi - \phi_{\rm R}$.
In the other $i$ - $\theta_{\rm R}$ range, the enhancement with the S-pole 
never show two sub-peaks, while that with the N-pole exhibits two sub-peaks 
in a range of $i$ as
\begin{equation}
\theta_{\rm R} - \phi_{\rm P} < i < \theta_{\rm R} + \phi_{\rm P},
\label{eqn:SP_Cond_i}
\end{equation}
calculated from equations (\ref{eqn:theta_LN}) and (\ref{eqn:thetaLN<phiP}).

Figure \ref{PulseProfileFigures} shows some examples of pulse profiles of emissions 
expected from the present model configurations 
for different combinations of $i$ and $\theta_{\rm R}$. 
Those of emissions from the polar cone and the polar mound 
are respectively presented in parallel there.
In a relatively soft X-ray band, say, in 2 - 5 kev, X-rays from the polar cone is dominant, 
while those from the polar mound is so 
in a relatively hard X-ray band, say, in 20 - 50 keV.
Thus, the left and right panels in figure \ref{PulseProfileFigures} can be considered to represent 
pulse profiles expected in the soft and hard band respectively. 

%%%%% figure {PulseProfileFigures} %%%%%

Pulse profiles observed from X-ray pulsars are classified into five groups
by Nagase (1989).
His classifications are shown, slightly changing the original wording, 
as follows, 
\begin{enumerate}
\item[(a)] single sinusoidal-like shapes both in the soft and hard bands,

\item[(b)] sinusoidal-like double peaks with little energy dependence, 
where the amplitudes of the two peaks are usually different,

\item[(c)] an asymmetric single peak with some features,

\item[(d)] a single sinusoidal-like peak in the high energy band and 
close adjacent double peaks in the soft energy band, and

\item[(e)] double sinusoidal-like peaks in the high energy band and complex five peaks 
in the soft energy band.
\end{enumerate}

Although sources on which the Nagase's classifications are based have a wide range of 
X-ray luminosities, we can identify sources with X-ray luminosities $\gtrsim 10^{37}$ erg s$^{-1}$ to groups (a) $\sim$ (d), as 
LMC X-4 (e.g. Hung et al. 2010) to (a), SMC X-1 (e.g. Raichur \& Paul 2010) to (b), Cen X-3 (e.g. Raichur \& Paul 2010) to (c) 
and Her X-1 (e.g. Deeter et al. 1998) to (d).  Vela X-1 (e.g. Makishima et al. 1999) is the representative source of the group (e) but its luminosity is as low as $\sim$ 
10$^{36}$ erg s$^{-1}$.

The present model can explain the basic features of the profiles (a) through (d), 
as seen from figure \ref{PulseProfileFigures}, where each pair of profiles (a) through (d) 
is shown as a typical example for the respective group (a) through (d).
Even for the group (e), we see presence of multiple peaks in the profile of the polar cone 
component in figure \ref{PulseProfileFigures} - (e), although the number of peaks 
is not five but four.

Even though the pulse profiles are mainly reproduced by adjusting two parameters, 
$\theta_{\rm R}$ and $i$, simultaneous adjusting of the other model parameters 
should be necessary 
in order to have the model profile fit to the observed profile as accurately as possible.

The beam width parameter, $\sigma_{\rm E}$, determines the width of the main peak 
in the pulse profile of the polar cone component.
Figure \ref{PLS_PRFL_ParameterDependence_1} shows the pulse profiles of the polar cone component when we change only the value 
of $\sigma_{\rm E}$ from 30$^{\circ}$ to 20$^{\circ}$ to 40$^{\circ}$ keeping 
the other parameters in the case of figure \ref{PulseProfileFigures} (c).
We can see that the peak profile gets significantly sharper with the $\sigma_{\rm E}$ 
decrease.
In rough comparisons of the models with the observations, 
$\sigma_{\rm E}$ around 30$^{\circ}$ seems appropriate.

%%%%% figure {PLS_PRFL_ParameterDependence_1} %%%%%

The relative distance, $\xi$, of the X-ray emitting position, normalized by the Schwarzshild radius of the neutron star mass, determines 
effects of the relativistic light-bending on pulse profiles.
One of the effects appears 
in depth of hollows between pulse peaks of the polar mound component.
Figure \ref{PLS_PRFL_ParameterDependence_2} shows respective pulse profiles of the polar mound component 
in cases of $\xi$ = 10 and 2.5 
when the other parameters are the same as those in figure \ref{PulseProfileFigures} - (e) in which $\xi$ = 5.
We see that the smaller $\xi$ makes the hollow shallower, suffering  
the larger light bending.

%%%%% figure {PLS_PRFL_ParameterDependence_2} %%%%%

The other effect produces small humps between main peaks in combination with 
the parameter, $\phi_{\rm C}$, which expresses the degree of the obscuration 
of X-rays by the neutron star body, defined in appendix.
Two examples are shown in figures \ref{PLS_PRFL_ParameterDependence_3}.
A low plateau begins to appear between two adjacent peaks when we change $\xi$ 
from 5 to 10 for the profile of the polar cone component in figure \ref{PulseProfileFigures} - (c), 
as in figure \ref{PLS_PRFL_ParameterDependence_3} - (A) - (left).  
It gets more significant if  we add a change of $\phi_{\rm C}$ from 60$^{\circ}$ to 30$^{\circ}$, 
as in figure \ref{PLS_PRFL_ParameterDependence_3} - (A) - (right).
This feature could explain the presence of a small hump only in the soft band 
in the observed pulse profile of Cen X-3 (see figure 12 in Raichur \& Paul 2009).

%%%%% figure {PLS_PRFL_ParameterDependence_3} %%%%%

If we change the combination of $\xi$ and $\phi_{\rm C}$ from (5, 60$^{\circ}$) 
to (10, 30$^{\circ}$), the pulse profile of the polar cone component in figure \ref{PulseProfileFigures} - (e) also converts to figure \ref{PLS_PRFL_ParameterDependence_3} - (B) - (left) and a small peak 
start to be exhibited.  
It gets higher when we further change $\phi_{\rm C}$ to 15$^{\circ}$ as seen in 
figure \ref{PLS_PRFL_ParameterDependence_3} - (B) - (right), where we can now produce six peaks in the model pulse profile.
 
As seen above, various combinations of the model-parameters can produce variety of 
pulse profiles but cannot form an asymmetry of a pulse-peak around the center of 
the peak.
Such asymmetries are often seen in observed pulse profiles and the typical example 
is the pulse profile of Cen X-3.
The similar asymmetry is seen in the pulse profile of Her X-1.
It is known, in case of Her X-1, that 
the pulse profile changes between the main-on phase and the mid-on phase 
of the 35-day on-off cycles.
Deeter et al. (1998) and Scott, Leahy and Wilson (2000) propose that 
the central X-ray emission regions are periodically obscured by the outer boundary of 
the magnetic funnels located at the inner edge of a precessing accretion disk, and that 
the pulse phase of the obscuration varies in association with geometry changes 
due to the disk precession.
By introducing such periodic obscuration by the outer boundaries of the magnetic 
funnels co-rotating with the central neutron star, we could interpret the asymmetries
seen in the pulse profiles from Cen X-3 and Her X-1 in the present model frame.

Obscuration could be done even by matter flowing along the magnetic funnels 
toward the neutron star surface, since optical depth for the electron scattering 
of the free fall region in the magnetic funnel is calculated in 
equation (\ref{eqn:tau}) to be larger than unity unless $r \gg 10^{7}$ cm.
In this case, the obscuration could be observed as an absorption dip in 
a pulse profile and be able to interpret some features in the complex multiple peaks 
as seen in the group (e).

%%%%%%%%%%%%%%%%%%%%%%%%%

\section{Summary and discussions}

Structures of X-ray emitting magnetic polar regions of neutron stars in X-ray pulsars are studied, 
and expected properties of X-ray emissions from them are compared with observations.

The matter flow in the polar cone is solved on such several assumptions as follow:
\begin{itemize}
\item The flow is one-dimensional in the radial direction, steady and sufficiently subsonic.
\item The opening angle of the cone is sufficiently smaller than unity 
and the physical quantities can be represented by respective typical values over the cross section of the cone.
\item The optical depth in the tangential direction of the cone is sufficiently large and 
the radiation energy density is greatly dominant to the gaseous one.
\item The energy loss of the flow is governed by photon diffusion in the tangential 
direction of the cone.
\end{itemize}

We discuss the validities of these assumptions below.

Equations (\ref{eqn:DifEq1}) and (\ref{eqn:DifEq2}) do not include 
the opening angle parameter, $\Theta$, 
and thus the solutions on the basic structure of the polar cone 
do not depend on the assumption of $\Theta$ in equation (\ref{eqn:Theta_r/R}).  
That assumption is necessary only to calculate the inflow velocity and  
the optical depth in the azimuthal direction.

As seen in figure \ref{Mdot-Height_Relations}, the height of the polar cone is close 
to the position of the magneto-boundary surface with the distance of $r_{\rm M,p}$, 
when $\dot{M}$ gets as large as several times 10$^{18}$ g s$^{-1}$.
Since $\Theta$ = 0.32 at $r_{\rm M,p}$ as obtained in equation  (\ref{eqn:Theta_0}), 
$\Theta \ll 1$ can be assured in the polar cone unless $\dot{M}$ is close to 
10$^{19}$ g s$^{-1}$.
The solved infall velocities are displayed in figures \ref{Solutions_B1} and \ref{Solutions_B10}, 
while the sound velocities in the polar cones should be 10$^{9} \sim 10^{10}$ cm s$^{-1}$. 
These confirm that the flow in the polar cone is sufficiently subsonic.
The sufficiently large optical 
depth can be recognized in figures \ref{Solutions_B1} and \ref{Solutions_B10}.

The radiation temperatures of the polar cone are plotted in figures \ref{Solutions_B1} and \ref{Solutions_B10}, 
while the gaseous temperature should be 10$^{11} \sim 10^{12}$ K 
in order for the gaseous energy density to comparable to the radiation energy density. 
Thus we see that the radiation energy is greatly dominant to the gaseous one 
in the polar cone.
The establishment of the radiation field just behind the shock front (the upper boundary)
has been discussed in \ref{RadEnDominance}.

The temperature at the bottom of the polar cone is simply calculated by the 
lower boundary condition as $P = B_{\rm r, *}^{2} / 8\pi$ to be 2.0 $\times 10^{9}$ 
for $B_{\rm r, *} = 10^{12}$ G, 
and the average energy of photons is $\sim$ 170 keV.
Since this energy is much larger than the cyclotron energy, $\sim$ 12 keV for 
$B_{\rm r, *} = 10^{12}$ G, the reduction of the electron-scattering opacity 
for photons passing along the magnetic lines force 
should not take place in the main part of the polar cone.  Thus, we can say that 
the simple geometry with $\Theta \ll 1$ determines the direction of the photon-diffusion, which should be the side-way direction.

Solutions show that the height of of the polar cone has a large dependence on the accretion rate.  
When $\dot{M} \simeq 10^{16}$ g s$^{-1}$, the height is a tenth as low as the neutron 
star radius.  On the other hand, when $\dot{M} \simeq 10^{18}$ g s$^{-1}$, the height 
is about 10 times as large as the neutron star radius.
This large dependence of the height on the accretion rate originates from 
the proportionality of $r_{\rm D}$ to the accretion rate, $\dot{M}$ as seen in 
equation (\ref{eqn:Def_rD}).
If we introduce the time scale for photons to diffuse out in the azimuthal direction, 
$t_{\rm D}$, it is given by replacing $r_{\rm S}$, $\rho_{\rm S}$ and $\Theta_{\rm S}$ 
in equation (\ref{eqn:tDS}) with $r$, $\rho$ and $\Theta$ 
for a general position in the polar cone.
Then, with the help of equation (\ref{eqn:ContEq}), the following relation is achieved:
\begin{equation}
r_{\rm D} = v \; t_{\rm D}.
\label{eqn:rD-tD_Rel}
\end{equation}
This means that $r_{\rm D}$ expresses the distance by which the infalling matter 
advances in the diffusion time.
Since the photon diffusion time is considered to be the cooling time of the radiation 
energy in the polar cone, $r_{\rm D}$ represents the cooling length in which 
the infalling matter significantly loses its energy unless a heating exists.
In fact, if we neglect the heating term, $GM/r^{2}$, in the right side of 
equation (\ref{eqn:DifEq1}), 
we get such a solution as an exponential decay of $\varepsilon$ 
with the scale length of (4/3) $r_{\rm D}$.
The rough proportionality of the height of the polar cone to $\dot{M}$ as seen in 
figure \ref{Mdot-Height_Relations} is basically explained by the proportionality 
of $r_{\rm D}$ to $\dot{M}$.

In practice, however, the term of $GM/r$ in the right side of equation (\ref{eqn:DifEq1}) 
cannot be neglected and rather plays the more important role 
with the larger accretion rate.
When $\dot{M}$ is as low as $10^{16}$ g s$^{-1}$, the energy carried with the free-fall 
matter is almost released soon after the shock.  However, when $\dot{M}$ is as large as $10^{18}$ g s$^{-1}$ or larger, the gravitational energy gain exceeds the energy 
loss due to the photon diffusion and a significant amount of energy piles up on  
the bottom side of the polar cone.  
In section \ref{Introduction} it has been mentioned that the presences of the two regions, 
the primary and secondary regions, were already introduced even in the mid of 1970's.
This paper quantitatively clarifies that X-ray emissions from the primary region is 
dominant when $\dot{M} \lesssim 10^{17}$ g s$^{-1}$, 
and that those from the secondary region becomes 
dominant when $\dot{M}$ exceeds 10$^{17}$ g s$^{-1}$.

As the accretion rate increases to 10$^{17} \sim 10^{18}$ g s$^{-1}$, 
the specific radiation energy remaining at the bottom of the polar cone increases.
Then, the radiation pressure should increase and exceed the magnetic pressure 
which holds the flow within the polar cone.  
As a result, the matter should expand in the tangential 
direction along the neutron star surface, dragging the magnetic lines of force, 
and form a mound-like structure (the polar mound region).
By assuming that the matter settles on the neutron star surface after losing its remaining energy through photon-diffusion in the polar mound 
and the excess radiation pressure 
balances with magnetic pressure enhanced by dragging, 
the structure of the polar mound  is simply 
calculated and the results are reasonable unless $\dot{M} \gg 10^{18}$ g s$^{-1}$.
Davidson (1973) originally introduced the concept of the mound 
as the optically thick region at the bottom of the magnetic funnel. 
It is here generalized to be composed of the polar cone and the polar mound, 
structures of which vary with the accretion rate.

From such configurations as discussed above, we can expect 
an X-ray spectrum composed of a multi-color black-body spectrum 
from the polar cone region and a quasi-single black-body spectrum 
from the polar mound region from X-ray pulsars 
with $\dot{M} \gtrsim 10^{17}$ g s$^{-1}$.  
These spectral properties basically agree with observations as discussed in \ref{X-ray_spectrum}.
For further confirmations of validities of the present model, the following two analyses are expected.
One is analyses of spectral variations linked with the pulse phase, as done by 
Kondo, Dotani \& Inoue (in preparation) for Her X-1. 
It could be able to 
resolve the overall spectrum of an X-ray pulsar into several components,  
which keep the constant spectral shapes but change their normalization factors 
with the pulse phase.
The other is to see a spectral change of a X-ray pulsar associated with the flux. 
The spectrum should get harder as the accretion rate increases, 
since the relative flux of the quasi-single blackbody component to the multi-color 
blackbody component increases with the accretion rate.

A fairly sharp pencil beam is expected together with a broad fan beam 
from the polar cone region, while a broad pencil beam from the polar mound region.
With these X-ray beam properties,  basic patterns in X-ray pulse profiles observed from 
X-ray pulsars can be explained too. 

The fairly sharp pencil beam is formed when the emitted photon energy is well lower than 
the cyclotron energy at the surface of the pencil beam.
The cyclotron energies observed from several X-ray pulsars are around 20 keV (e.g. Makishima et al. 2000), while the typical X-ray energies from the polar cone are around 
a few keV.  This confirms the above condition for the fairly sharp pencil beam to appear.
If we see the softer X-rays from an X-ray emitting place in the polar cone far from the stellar surface, 
however, the cyclotron energy rapidly decreases in proportion to $r^{-3}$ while 
the surface temperature does not so largely.
Hence, the beaming effect should tend to weaken in the spectral range below 1 keV.

We have a number of freedoms in the present phenomenological model for 
the pulse profiles to make fine tuning the model pulse profile to reproduce  
an observed pulse profile.  
In particular, there could be large uncertainties in the periodic obscuration by 
the outer-boundary and the lower side of the free fall region of the magnetic funnel. 
The magnetic funnel should be bent in the rotational direction around the 
axis of the total angular momentum of the accreted matter at the magneto-boundary surface, since the angular momentum should be transferred though the magnetic stress 
to the neutron star within the magneto-boundary surface.
The degree of the bend of the magnetic lines of force depends 
on the amount of the angular momentum to be 
transfered (see e.g. Inoue 1976).
Thus, if the amount of the angular momentum changes, the pulse phase at which 
obscuration by the magnetic funnel takes place should change.
In fact, Her X-1 exhibits the pulse-profile change linked with the 35 day on-off cycle.
The super-orbital modulation of the X-ray flux can well be explained 
by the precession of the accretion disk (Inoue 2019).  
If so, the angular momentum axis of the accreted 
matter changes with the precession and the periodic phase shift 
of the obscuration place in the pulse profile could be understandable.
It could be possible to resolve the practical configurations of the magnetic funnel
by adjusting the model pulse profile to an observed profile.

This model can explain basic features in the pulse-profiles of many
X-ray pulsars as discussed in \ref{Expected_Pulse_Profiles} 
but it has not been done yet to show a good evidence 
for the present model to reproduce an observed pulse profile precisely.  
It should be done in future.\\

In this paper, we have basically dealt with cases of the accretion rate in the range of 
10$^{17} \sim 10^{18}$ g s$^{-1}$.
In this accretion rate range, we have seen that the fairly tall polar cone stands 
and the fairly wide polar mound extends from the bottom of the polar cone 
on each of the magnetic polar region of the neutron star.
The strength of the magnetic field of the neutron star has been assumed to be 
10$^{12} \sim 10^{13}$ gauss at its surface and the obtained configurations of the X-ray emitting polar region do not largely depend on it around this magnetic field range.

When the accretion rate is as low as 10$^{16}$ g s$^{-1}$ or less, 
the height of the polar cone becomes as low as or lower than a tenth of 
the neutron star radius.
Thus, such cases do not fit to the main assumption of the present study  
that the photon diffusion in the tangential direction plays the main role to construct 
the polar cone structure.
Several studies have, however, been being done on this low accretion rate case 
by several authors (e.g.  Becker \& Wolff 2007, Wolff et al. 2016, and references therein) 
since the pioneer works done by Davidson (1973) and Basko \& Sunyaev (1975).

When the accretion rate is as high as 10$^{19}$ g s$^{-1}$ or higher, on the other hand,
the luminosity largely exceeds the Eddington luminosity of 1 $M_{\odot}$ star.
Although the solution can exist in the present scheme of the magnetic polar regions 
even in the case of $\dot{M} \simeq 10^{19}$ g s$^{-1}$, 
some obtained parameters are inconsistent with the approximations in this study.

The first issue is on the base radius of the polar mound, $x_{\rm PM}$, as shown in figure \ref{MoundParameters}.  
We see that it exceeds the stellar radius when $\dot{M}$ gets close 
to 10$^{19}$ g s$^{-1}$.
Although this is the result from the very simplified treatment, 
the tendency could be understandable for the polar mound to extend over the entire surface of the neutron star 
as the accretion rate increases.  Taking account of it that the portion of the energy 
loss rate from the polar mound gets close to unity when $\dot{M}$ gets close 
to 10$^{19}$ g s$^{-1}$, as seen in figure \ref{EnergyLossPortions}, 
most of the super-Eddington luminosity should quasi-spherically be emitted 
from the entire surface of the neutron star with this accretion rate. 
It could, then, be difficult for the polar mound to maintain the steady state 
under the super-Eddington radiation pressure. 
The effect of the radiation pressure might, however, be weakened 
due to the reduction of the Thomson scattering cross section 
under the strong magnetic field on the surface of the neutron star. 

The second is on the height of polar cone.  It becomes as high as or higher than that of 
the magneto boundary surface when $\dot{M} \gtrsim 10^{19}$ g s$^{-1}$  
as shown in figure \ref{Mdot-Height_Relations}.  
If the upper boundary of the polar cone reaches 
the magneto-boundary surface, the matter trying to flow into the magnetic funnel 
should flood over the magneto-boundary surface. 
In that case, the approximations around the upper boundary of the polar cone in 
the present study become inapplicable and we will need some new treatments on them.

In addition to them, the optical depth of the accreted matter surrounding the central 
X-ray source could be a problem.
The optical depth of the transition layer of the magneto-boundary surface, 
$\tau_{\rm T}$, is roughly calculated from equation (\ref{eqn:CntEq_TL}) as 
\begin{eqnarray}
\tau_{\rm T} &\simeq& \kappa_{\rm T} \rho_{\rm T} h_{\rm T} \nonumber \\ &\simeq& \frac{\dot{M}}{4\pi r v_{\rm T} \sin \theta} \nonumber \\
&=& 3.3 \; \beta^{-1} \sin^{-1} \theta \left(\frac{r}{r_{\rm M,e}}\right)^{-1/2} \left(\frac{\dot{M}}{10^{19}\; \rm{g \; s}^{-1}}\right)^{8/7} \left(\frac{M}{M_{\odot}}\right)^{-1/14} \left(\frac{\mu_{\rm M}}{10^{30} \rm{gauss cm}^{3}}\right)^{2/7},
\label{eqn:tau_M}
\end{eqnarray}
where we have expressed $v_{\rm T} = \beta (2 GM/r)^{1/2}$.
Since $\beta$ must be significantly smaller than unity, 
the matter in the transition layer is considered to be 
optically thick everywhere in the case of $\dot{M} \gtrsim 10^{19}$ g s$^{-1}$.
If so, the super-Eddington radiation pressure should interfere with steady accretion.

This estimation is, however, on the assumption that the magnetic axis aligns with the 
rotational axis of the neutron star and both axes are perpendicular to the equatorial 
plane of the accretion disk.
In practical cases of X-ray pulsars, the magnetic axis must be oblique to the rotational 
axes, and thus the matter distribution in the transition layer should be biased to 
the equatorial plane of the disk.
In those cases, a part of the transition layer could be optically thin and steady accretion, 
in which the matter inflow and the photon outflow segregate from each other, might be possible even if the luminosity exceeds the Eddington limit.

These issues in the case of $\dot{M} \gtrsim 10^{19}$ g s$^{-1}$ as discussed above 
are quite interesting in relation to 
ultra-luminous X-ray pulsars recently discovered one after 
another (Bachetti et al. 2014; F$\ddot{\rm u}$rst et al. 2016; Israel et al. 2017; Carpano et al. 2018), and wait for future studies.\\

This study presents several clues to resolve 
observational evidences of X-ray pulsars from various viewpoints, but 
is still based on several simplified assumptions and approximations.
Further theoretical and observational studies of X-ray pulsars are largely expected.

\bigskip

\begin{ack}
The author greatly appreciates the referee for his or her critical comment on the original manuscript.
\end{ack}

\appendix

\section{Deformations of pulse profiles by the central neutron star}
We discuss two effects deforming the pulse profile of X-ray pulsars 
owing to the central neutron star.

One is an occultation of X-rays emitted from the polar cone by the neutron star body. 
When the line of sight of those X-rays from the polar cone 
crosses the neutron star body, they cannot be observed. 
A critical direction within which emitted X-rays suffer from this effect 
depends on the height of the X-ray emitting position on the polar cone, 
but we simultaneously observe X-rays from the polar cone in a fairly wide range 
of the height.
Thus, we introduce the following deformation factor, $A$,  
in calculating expected pulse profiles as
\begin{equation}
A =  \left\{
\begin{array} {ll}
 1 & \rm{when} \; \; \phi \leq \pi/2 \\
\exp \left (- \left[\frac{(\phi - \pi/2)}{\phi_{\rm c}} \right]^{2} \right) & \rm{when} \; \; \phi > \pi/2,
\end{array}
\right.
\label{eqn:F_cf_obs}
\end{equation}
where $\phi_{\rm C}$ is the average critical angle over the region of the polar cone  
emitting relevant X-rays.

Another is the relativistic light bending effect by the gravitational field of the neutron star.
To include this effect, we adopt the simple analytic formula to describe it which was 
introduced by Leahy \& Li (1995).
The applicable range of their original formula is limited in $2 \leq \xi \leq 5$, 
supposing the photon emitting place is on the surface of the neutron star.
Here, $\xi$ is a relative distance of a photon emitting place to 
the Schwarzschild radius.
In the present study, our interests are also in emissions from fairly far places of 
the polar cone from the neutron star surface and thus we have expanded the applicable range of $\xi$ up to $\sim 10$ or more by recalculating the necessary parameters 
in the same way as used in Leahy \& Li (1995).
The formula we have used is as follows.
\begin{equation}
\cos \phi' = a \cos \phi + b, 
\label{eqn:LightBend}
\end{equation}
where
\begin{equation}
a = \frac{2.304}{\xi^{2}} + \frac{0.653}{\xi} + 1.017,
\label{eqn:Prm_a}
\end{equation}
\begin{equation}
b = -\frac{2.490}{\xi^{2}} - \frac{0.584}{\xi} - 0.021.
\label{eqn:Prm_a}
\end{equation}
Here, $\phi$ is an angle, at which the light is emitted from a place with a 
relative distance, $\xi$, to the radial direction from the gravity center, 
and $\phi'$ is an angle at which the light is observed at infinity,  
to the original radial direction.

%In the theory (\cite{key-1})..........

%\subsection{Subsection}

%In the theory (\cite{key-2})..........

%\subsubsection{Subsubsection}

%The resent result from ...........

% See the manual for the detail.
%%%

\end{document}